\documentclass[twocolumn,amsmath,amssymb,nofootinbib,pra,superscriptaddress]{revtex4}

\usepackage{graphicx} 
\usepackage{epstopdf} 
\usepackage{pdfpages} 
\usepackage{microtype} 
\usepackage{booktabs} 
\usepackage{tensor} 

\begin{document}

\title{A variational approach to repulsively interacting three-fermion
  systems in a one-dimensional harmonic trap}
\date{\today}

\author{N.~J.~S. \surname{Loft}}
\affiliation{Department of Physics and Astronomy, Aarhus University, DK-8000 Aarhus C,  Denmark}
\author{A.~S. \surname{Dehkharghani}}
\affiliation{Department of Physics and Astronomy, Aarhus University, DK-8000 Aarhus C,  Denmark}
\author{N.~P. \surname{Mehta}}
\affiliation{Department of Physics and Astronomy, Trinity University, San Antonio, Texas, USA}
\author{A.~G. \surname{Volosniev}}
\affiliation{Institut f{\"u}r Kernphysik, Technische Universit{\"a}t Darmstadt, 64289 Darmstadt, Germany}
\affiliation{Department of Physics and Astronomy, Aarhus University, DK-8000 Aarhus C,  Denmark}
\author{N.~T. \surname{Zinner}}
\affiliation{Department of Physics and Astronomy, Aarhus University, DK-8000 Aarhus C,  Denmark}

\begin{abstract}
  We study a three-body system with zero-range interactions in a
  one-dimensional harmonic trap. The system consists of two
  spin-polarized fermions and a third particle which is distinct from
  two others (2+1 system).  First we assume that the particles have
  equal masses. For this case the system in the strongly and weakly
  interacting limits can be accurately described using wave function
  factorized in hypercylindrical coordinates. Inspired by this result we
  propose an interpolation ansatz for the wave function for arbitrary
  repulsive zero-range interactions. By comparison to numerical
  calculations, we show that this interpolation scheme yields an
  extremely good approximation to the numerically exact solution both
  in terms of the energies and also in the spin-resolved densities.
  As an outlook, we discuss the case of mass imbalanced systems in the
  strongly interacting limit. Here we find spectra that demonstrate
  that the triply degenerate spectrum at infinite coupling strength of
  the equal mass case is in some sense a singular case as this
  degeneracy will be broken down to a doubly degenerate or
  non-degenerate ground state by any small mass imbalance.
\end{abstract}

\maketitle


\section{Introduction}
Recent advances in cold atomic gas experiments has made it possible to
work with microscopic system sizes for fermionic
\cite{serwane2011,zurn2012,wenz2013,zurn2013} and bosonic samples
\cite{will2011,nogrette2014,labuhn2014,will2014}. Furthermore, by
application of optical lattices \cite{bloch2008} and use of Feshbach
resonances \cite{chin2010} it is possible to tune both the geometry
and the interaction strength of these setups. This allows the cold
atom systems to address a host of interesting physical models in lower
spatial dimensions that are typically not so easily accessible in
other fields. In particular, when the system is squeezed down to a
regime where particles effectively move along just a single spatial
direction, one can hope to realize some of the exactly solvable models
that are known for both few- and many-body systems in one dimension
(1D) \cite{sutherland2004,cazalilla2011}. About a decade ago this hope
led to the realization of the strongly repulsive (hard-core) Bose gas
\cite{paredes2004,kinoshita2004,kinoshita2005} in the so-called
Tonks-Girardeau regime \cite{tonks1936,girardeau1960,olshanii1998} and
later on also to the so-called super-Tonks-Girardeau gas \cite{astrak2005} 
which is an excited state for strong attractive interactions
\cite{haller2009}. More recently, the strongly repulsive and
attractive regimes have been explored with few-body systems of
two-component fermions \cite{zurn2012,wenz2013,zurn2013} and these
recent experimental developments provide a major motivation for the
current work.

The experimental progress has generated great interest for few-body
problems in one-dimensional geometries for both bosonic
\cite{zollner2005,deuret2007,tempfli2008,girardeau2011,brouzos2012,brouzos2014,wilson2014,zinner2014},
fermionic
\cite{guan2009,yang2009,girardeau2010,guan2010,rubeni2012,astrak2013,brouzos2013,bugnion2013,gharashi2013,sowinski2013,volosniev2013,lindgren2014,gharashi2014,deuret2014,volosniev2014,cui2014,sowinski2014,levinsen2014},
and mixed systems
\cite{girardeau2004,girardeau2007,zollner2008,deuret2008,fang2011,harshman12,garcia2013a,garcia2013b,harshman2014,campbell2014,garcia2014a,damico2014,mehta2014,dehk2014,garcia2014b}.
Recently, it has been shown that for strong short-range repulsive
interactions a 1D two-component Fermi system in a harmonic trap
exhibits strong magnetic correlations already at the three-body level
\cite{gharashi2013,lindgren2014}.  More generally, one finds that in
the ground state of strongly interacting $N+1$ system the impurity
will be mainly observed in the middle of the trap
\cite{lindgren2014,levinsen2014}. This result can be generalized to
other types of 1D confinement \cite{volosniev2013,deuret2014}.  This
should be contrasted to two-component bosonic systems with equal
strength intra- and interspecies interactions where the ground state
for strong repulsion will be the one predicted by Girardeau
\cite{girardeau2011} and the impurity would be essentially delocalized
\cite{volosniev2013}.  In the present paper we seek further analytical
and semi-analytic insights into the 1D fermionic three-body problem in
a harmonic trap by constructing a class of variational wave functions
for arbitrary repulsive zero-range interaction strength. Relying on
our knowledge developed for the weakly and strongly interacting limits
we provide and study a variational wave function of the three-body
problem that connects these two limits. By comparison to numerical
results we show that our class of states yields an exceptionally good
approximation for the low-energy part of the energy spectrum and also
gives very accurate spin-resolved densities. This shows that intuitive
approaches at the level of the wave function shape are effective in
strongly interacting 1D few-fermion systems. For related recent work
on single-component bosons see Refs.~\cite{brouzos2012,wilson2014} and 
for recent work on the two-component three-body bosonic system see 
Ref.~\cite{garcia2014b}.  As an outlook
we consider the 2+1 system in the case where the masses are imbalanced
and find an intriguing change in the ground state structure for strong
interactions which occurs for any infinitesimal difference in the
masses between the two components. Our results indicate that the large
degeneracy of strongly interacting two-component systems is in some
sense accidental and that spectrum for equal masses is in fact a
special case (although of course an extremely important one).

The paper is organized as follows. In section~\ref{sec:system} we
introduce the system, our choice of coordinates and discuss
the symmetries of our Hamiltonian.
In section~\ref{sec:limits} 
we solve the problem for zero and infinite zero-range interaction
strength while the variational approach to arbitrary repulsive
interaction strength is discussed in Section~\ref{sec:variation}. In
Section~\ref{sec:masses} we provide an outlook towards the case where
the two components have unequal masses by solving the general problem
in the strongly interacting regime. Section~\ref{sec:conclusions}
contains our conclusions and outlook. Finally, we provide three
appendices with technical details of important derivations discussed
in the main text.


\section{The system}
\label{sec:system}

In this section, we introduce the system that will be the subject for
the rest of the article. The section is largely based on
reference~\cite{harshman12} and is mainly concerned with different
coordinate systems in which the system can be described.  At the end
of this section we discuss the parity and permutation symmetries of
the Hamiltonian.

\subsection{The Hamiltonian and coordinate transformations}

Consider the Hamiltonian for a system of $N$ particles in one
dimension
\begin{equation*}
   H =  H_0 +  V
\end{equation*}
\noindent consisting of a harmonic trap Hamiltonian
\begin{equation}
  \label{eq:2}
   H_0 = \sum\limits_{i=1}^N
  \left( \frac{{\tilde p_i}^2}{2m_i} +
      \frac{m_i\,\omega^2}{2} {\tilde q_i}^2 \right)
\end{equation}
\noindent and an interaction term
\begin{equation}
  \label{eq:3}
   V = \sum\limits_{i=1}^{N-1} \, \sum\limits_{i<j}^{N} \,
  g_{ij} \, \delta \! \left(\tilde q_i - \tilde q_j\right) \; ,
\end{equation}
\noindent where $\tilde q_i$ and $\tilde p_i = -i\hbar (\partial
/ \partial \tilde q_i)$ are correspondingly the position and momentum
operators of particle $i$, and $m_i$ is its mass. The first part
\eqref{eq:2} describes $N$ such particles in a harmonic oscillator
potential with angular frequency $\omega$. The second
part~\eqref{eq:3}, containing Dirac's delta functions, describes a
contact interaction between particles $i$ and $j$ of strength
$g_{ij}$.

In this article everywhere except section~\ref{sec:masses}, we shall
limit ourselves to $N=3$, $m_1 = m_2 = m_3 \equiv m$, and $g_{13} =
g_{23} \equiv g \geq 0$. That is, we first consider a system of three
particles of equal mass. We take particles 1 and 2 to be spinless
(spin-polarized) fermions interacting with the third particle with
strength $g$.  Due to the Pauli principle the wave function should
vanish whenever particles 1 and 2 meet, thus the corresponding
contribution from the delta function interaction should be
neglected. Having this in mind we assume $g_{12}=0$ to simplify
notation.

\begin{figure}[t]
  \centering
  \includegraphics[width=0.7\columnwidth]{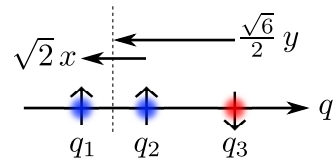}
  \caption{The coordinates $x$ and $y$ describe the relative position
    of the particles as opposed to $q_1$, $q_2$ and $q_3$ describing
    the absolute positions. The illustration shows the scaled
    coordinates $\sqrt{2} \, x = q_1 - q_2$ and $\tfrac{\sqrt{6}}{2}
    \, y = \tfrac{1}{2}\left(q_1 + q_2 \right) - q_3$, the latter
    pointing from $q_3$ to the center-of-mass coordinate of the
    particle 1 and 2 subsystem. Particle 1 and
    2 are indistinguishable, but distinguishable from particle 3, like
    for instance two spin-up particles vs. one spin-down.}
  \label{fig:trepartikler}
\end{figure}

Also, we introduce the length scale $\sigma = \sqrt{\hbar / m\omega}$
such that more convenient dimensionless coordinates can be defined as
\begin{equation*}
  q_i = \frac{\tilde q_i}{\sigma}
  \quad \text{and} \quad
  p_i = \frac{\tilde p_i \, \sigma}{\hbar} \; .
\end{equation*}
\noindent In these coordinates the Hamiltonian of the three particle
system becomes
\begin{equation}
  \label{eq:H0_car}
   H_0 = \frac{\hbar\omega}{2}
  \sum\limits_{i=1}^3 \left({p_i}^2 + {q_i}^2\right)\; ,
\end{equation}
\begin{equation}
  \label{eq:V_car}
   V = \frac{g}{\sigma} \, \delta \! \left(q_1 - q_3\right)
  + \frac{g}{\sigma} \, \delta \! \left(q_2 - q_3\right) \; .
\end{equation}
\noindent Now we choose units such that $\hbar \equiv \omega \equiv m
\equiv 1$, then it follows that also $\sigma=1$. It is possible to
separate the center-of-mass motion and the relative motion of the
particles if we define a new set of coordinates $\mathbf{r} \equiv
[x,y,z]^{T}$ by applying a linear transformation to $\mathbf{q}
\equiv[q_1,q_2,q_3]^T$ given by the matrix $\mathbf{J}$, that is
\begin{equation}
  \label{eq:jacobian}
  \mathbf{r} = \mathbf{J} \mathbf{q}
  \quad \text{or} \quad
  \begin{bmatrix} x \\[0.5em]
                  y \\[0.5em]
                  z \end{bmatrix} =
  \begin{bmatrix} \frac{1}{\sqrt{2}} &
    -\frac{1}{\sqrt{2}} & 0 \\[0.5em]
    \frac{1}{\sqrt{6}} & \frac{1}{\sqrt{6}} &
    -\frac{\sqrt{2}}{\sqrt{3}} \\[0.5em]
    \frac{1}{\sqrt{3}} & \frac{1}{\sqrt{3}} &
    \frac{1}{\sqrt{3}} \end{bmatrix}
    \begin{bmatrix} q_1 \\[0.5em]
                    q_2 \\[0.5em]
                    q_3 \end{bmatrix} \; .
\end{equation}
\noindent The new coordinates, $(x,y,z)$, are called the standard
normalized Jacobi coordinates. Since $\mathbf{J}^T\mathbf{J} =
\mathbf{1}$ is the identity matrix and $\det\mathbf{J} = 1$, the
matrix $\mathbf{J}$ is a member of the three dimensional rotation
group $\text{SO}(3)$. Therefore $\mathbf{r}$ is merely a rotation of
$\mathbf{q}$ and the norm of the vector is conserved,
i.e. $\mathbf{r}^2 = \mathbf{q}^2$. Since $\mathbf{J}^{-1} =
\mathbf{J}^T$, the inverse relation is given by $\mathbf{q} =
\mathbf{J}^T \mathbf{r}$. As is seen from eq.~\eqref{eq:jacobian}, $z$
describes the center-of-mass position of the system since all
individual positions of the particles are weighted equally. The
relative motion is described with coordinates $x$ and $y$, as
visualized in figure~\ref{fig:trepartikler}.

Similarly, we rotate the momenta coordinates, $\mathbf{p}$, such that
$\mathbf{k} = \mathbf{J}\mathbf{p}$, where $\mathbf{k}^2 =
-{\mbox{\boldmath $\nabla$}_\mathbf{r}}^2$ -- the subscript
$\mathbf{r}$ denotes that the differentiation is done with respect to
the $\mathbf{r}$-coordinate system. In the Jacobi coordinates, the two
terms of the Hamiltonian become
\begin{align}
  \label{eq:H0_jacobi}
   H_0 &= \frac{1}{2} \left(\mathbf{r}^2 -
     {\mbox{\boldmath $\nabla$}_\mathbf{r}}^2\right)
   \nonumber\\[0.3em]
   &= \frac{1}{2} \left(
    x^2 - \frac{\partial^2}{\partial x^2} + y^2 -
    \frac{\partial^2}{\partial y^2} + z^2 -
    \frac{\partial^2}{\partial z^2} \right) \; ,
\end{align}
\begin{equation}
  \label{eq:V_jacobi}
   V = \frac{g}{\sqrt{2}} \,
          \Bigg[ \delta \! \left(-\tfrac{1}{2}x + \tfrac{\sqrt{3}}{2}y\right) +
          \delta \! \left(-\tfrac{1}{2}x - \tfrac{\sqrt{3}}{2}y \right) \Bigg] \; .
\end{equation}
\noindent Notice that $ H_0$ is identical to the Hamiltonian for a
single particle at position $\mathbf{r}$ in a three dimensional
harmonic oscillator. It is clearly separable in all of its
coordinates, and each term has the well-known energy eigenbasis of a
one dimensional harmonic oscillator. However, the interaction term,
$V$, is not separable in its coordinates, fortunately it only depends
on $x$ and $y$, so the total Hamiltonian, $ H$, can be separated in
terms of the center-of-mass motion ($z$-direction) and relative motion
($xy$-plane). Since the relative motion of the particles belongs to
the $xy$-plane, we define one last set of coordinates to get the most
beneficial description of this plane:
\begin{align*}
  &\rho = \sqrt{x^2 + y^2} \; , \quad \rho \in [0,\infty[ \; ; \\[0.5em]
  &\tan\phi = \frac{y}{x} \; , \quad \phi \in [-\pi,\pi[ \; .
\end{align*}
\noindent The set $(\rho,\phi,z)$ is called the Jacobi
hypercylindrical coordinates. The trap potential and the interaction
potential take the form
\begin{equation}
  \label{eq:H0_cyl}
   H_0 = \frac{1}{2} \left(z^2 - \frac{\partial^2}{\partial z^2}
  + \rho^2 -\frac{1}{\rho}\frac{\partial}{\partial \rho}
  - \frac{\partial^2}{\partial \rho^2}
  - \frac{1}{\rho^2} \frac{\partial^2}{\partial \phi^2} \right) \; ,
\end{equation}
\begin{equation}
  \label{eq:V_cyl}
   V = \frac{g}{\sqrt{2}\,\rho} \Big[
      \delta \! \left(\phi - \tfrac{\pi}{6}\right)
    + \delta \! \left(\phi + \tfrac{5\pi}{6} \right)
    + \delta \! \left(\phi + \tfrac{\pi}{6}\right)
    + \delta \! \left(\phi - \tfrac{5\pi}{6}\right)
    \Big] \; .
\end{equation}

\begin{figure}[htbp]
  \centering
  \includegraphics[width=\columnwidth]{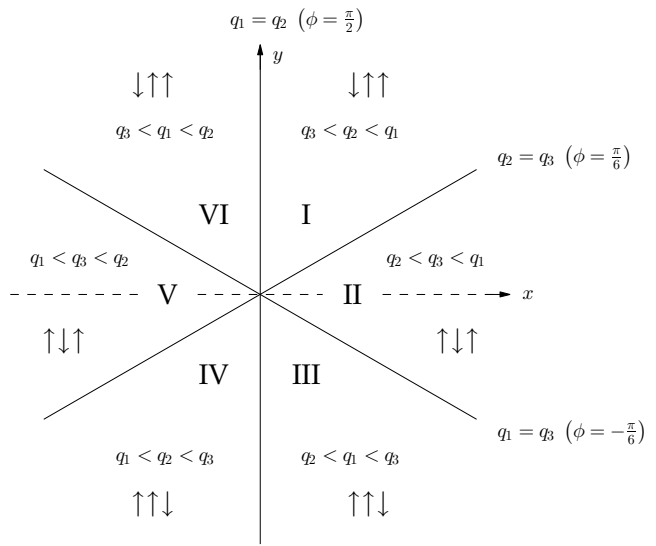}
  \caption{The relative configuration space showing all possible ways
    to order the three particles relative to each other. Thinking of
    spin $\tfrac{1}{2}$ particles, we could take particle 1 and 2
    to be in the spin-up state and particle 3 to be in the spin-down
    state, and these spin configurations are also depicted.}
  \label{fig:hypersfaerisk}
\end{figure}

In figure \ref{fig:hypersfaerisk} we show the relative configuration
space for the particles, which will become helpful when we describe
the wave function in the following sections. By relative configuration
space we mean that any relative configuration of the three particles
is uniquely determined by a single point in the plane, this point
being given in either $(x,y)$ or $(\rho,\phi)$ coordinates. To include
all absolute configurations, we would need to include a third
dimension, namely the $z$-axis, since this determines the
center-of-mass position. The solid lines on the figure represent two
particles sharing the same position. Since we assume a contact
interaction between two distinguishable particles the delta functions
in eq.~\eqref{eq:V_cyl} are non-zero only on the solid lines $q_2 =
q_3$ and $q_1 = q_3$.

\subsection{Parity and permutation symmetry}

From \eqref{eq:H0_car} and \eqref{eq:V_car} we see that the total
Hamiltonian is invariant under a simultaneous change of sign in all
spatial coordinates. If $ \Pi$ denotes the parity operator
transforming $q_i \mapsto -q_i$ for every $i \in \{ 1,2,3 \}$, then
surely $\left[ H , \Pi\right] = 0$. In the Jacobi coordinates the
parity transformation is $(x,y,z) \mapsto (-x,-y,-z)$, and so the
parity operator can be decomposed as $ \Pi = \Pi_{xy} + \Pi_z$, where
the first term acts only on coordinates in the relative $xy$-plane and
the last term acts only the coordinate on the $z$-axis. We choose this
decomposition because the Hamiltonian is separable in the same
way. From \eqref{eq:H0_jacobi} and \eqref{eq:V_jacobi} $\left[ H ,
  \Pi_z\right] = 0$, and then $\left[ H , \Pi\right] = \left[ H ,
  \Pi_{xy}\right] + \left[ H , \Pi_z\right] = \left[ H ,
  \Pi_{xy}\right]=0$. From now on the term \emph{parity} will refer to $
\Pi_{xy}$.

By construction the Hamiltonian is invariant under the exchange of
particles 1 and 2. It immediately follows that $\left[ H ,
  P_{12}\right] = 0$ where $ P_{12}$ denotes the permutation operator
exchanging the coordinates of particles 1 and 2. Since the
transformation $x \mapsto - x$ is equivalent to $q_1 \leftrightarrow
q_2$, the Pauli principle is satisfied if and only if the wave
function fulfills $\psi(-x,y,z) = -\psi(x,y,z)$. One obvious
consequence is that the wave function must vanish on the $y$-axis in
figure \ref{fig:hypersfaerisk}.

We easily verify that $\left[ \Pi_{xy} , P_{12}\right] = 0$, and that
$ H$, $ \Pi_{xy}$ and $ P_{12}$ are all hermitian. Therefore we may
find a basis that is simultaneously described by energy, parity and
permutation of particles 1 and 2. The eigenvalues for both the parity
operator and the permutation operator are $\pm 1$. However, the Pauli
principle states that only eigenfunctions with the eigenvalue $-1$ for
$ P_{12}$ are valid wave functions (the eigenvalue $+1$ is only for
bosonic wave functions as discussed in Refs.~\cite{harshman12,garcia2014b}). We say that the states with parity eigenvalue
$+1$ have even parity, and the states with $-1$ have odd parity. Thus
we require
\begin{align*}
  &\psi(-x,y,z) = -   \psi(x,y,z)  &&\text{(Pauli principle)}\\
  &\psi(-x,-y,z) = \pm \psi(x,y,z)   &&\text{(even/odd parity)}
\end{align*}


\section{The interaction limits}
\label{sec:limits}

In this section, we find the exact wave functions that solve the
Schr\"odinger equation at $g=0$ and $1/g=0$.

\subsection{Non-interacting limit, $g=0$}

Without the interaction the system can be considered as a
three-dimensional quantum harmonic oscillator with Hamiltonian
$H_0$. Here we write down the eigenspectrum of this textbook
Hamiltonian using $(\rho,\phi,z)$ coordinates. If we denote an energy
eigenbasis of $H_0$ in this set of coordinates by $\left| \nu,\mu,\eta
\right>$, we separate the center-of-mass motion and the relative
motion as $\left| \nu,\mu,\eta \right> = \left| \nu,\mu \right>
\otimes \left| \eta \right>$. The wave functions for the
center-of-mass motion are the eigenstates of the one dimensional
harmonic oscillator, i.e.
\begin{align}
  \label{eq:cm_wf}
  \left< z \, \right| \left. \! \eta \right>
  &= \psi_\eta(z) \nonumber\\
  &= \frac{\pi^{-1/4}}{\sqrt{2^\eta \eta !}}
  \, e^{-z^2 / 2}\,
  H_\eta(z) \; , \quad \eta = 0, 1, 2, \dots
\end{align}
where $H_\eta(z)$ denotes the Hermite polynomial of degree
$\eta$. The relative motion is described with
functions\footnote{Due to the Pauli principle it is impossible to have
  $\mu = 0$.}
\begin{align}
  \label{eq:relative_wf}
  \left< \rho,\phi \right| \left. \! \nu,\mu \right> &=
  \psi_{\nu,\mu}(\rho,\phi) \nonumber\\
  &= A \cdot
  L_\nu^{(\mu)} \! \big(\rho^2\big) \, e^{-\rho^2 / 2} \, \rho^\mu \,
  f(\mu,\phi) \; , \\
  &\; \nu = 0, 1, 2, \dots
  \quad \text{and} \quad \mu = 1, 2, 3 \dots  \; ,
  \nonumber
\end{align}
\noindent where $A$ is a normalization constant, $L_\nu^{(\mu)} \!
\big(\rho^2\big)$ denotes the associated Laguerre polynomial and
$f(\mu,\phi)$ contains the angular dependency of the wave function and
is in the simultaneous energy and parity eigenbasis either equal to
$\sin(\mu\phi)$ or $\cos(\mu\phi)$ (see below). However, we keep the
general notation for the angular function $f(\mu,\phi)$. Since
$L_\nu^{(\mu)} \! \big(\rho^2\big)$ have $\nu$ roots, this is the
quantum number determining the number of roots in the radial
$\rho$-direction and thus we refer to $\nu$ as a radial excitation
quantum number. Also we note that the value of $\mu$ determines the
number of roots in the angular $\phi$-direction, and so we regard
$\mu$ as an angular excitation quantum number. The quantum number for
the center-of-mass excitation is $\eta$.  The energy corresponding to
these quantum numbers is given as
\begin{equation}
  \label{eq:energy}
  E = \left< \nu,\mu,\eta \vphantom{H_0}
    \right| H_0 \left| \nu,\mu,\eta  \vphantom{H_0} \right>
    = \tfrac{3}{2} + 2\nu + \mu + \eta \; .
\end{equation}

\subsection{Impenetrable regime, $1/g=0$}
\label{sec:impenetrable}

As discussed in the previous section, the center-of-mass motion is
separable for all values of $g$, and so we will focus on the wave
function describing the relative motion. To start the discussion we
first derive conditions for the wave function on the lines of
interaction in figure~\ref{fig:hypersfaerisk}. Let $\phi_0 \in \big\{
\pm \tfrac{\pi}{6}, \pm \tfrac{5\pi}{6} \big\}$, we then integrate the
time-independent Schr\"odinger equation, $H\psi=E\psi$, in the
$\varepsilon$-neighborhood of $\phi_0$ and let $\varepsilon
\rightarrow 0$:
\begin{align*}
  &  -\frac{1}{2\rho^2}
  \lim\limits_{\varepsilon \rightarrow 0}
  \int\limits_{\phi_0-\varepsilon}^{\phi_0+\varepsilon} \mathrm{d}\phi \,
  \frac{\partial^2}{\partial\phi^2} \psi(\rho,\phi) \\
  & + \frac{g}{\sqrt{2}\, \rho}
  \lim\limits_{\varepsilon \rightarrow 0}
  \int\limits_{\phi_0-\varepsilon}^{\phi_0+\varepsilon} \mathrm{d}\phi \,
  \delta \! \left(\phi_0 - \phi\right) \, \psi(\rho,\phi) =0
\end{align*}
\noindent All other terms vanish due to the continuity of the wave
function. The remaining integrals yield
\begin{equation*}
  -\frac{1}{2\rho^2} \lim\limits_{\varepsilon \rightarrow 0}
  \left( \left. \frac{\partial\psi}{\partial\phi}
    \right|_{\phi_0 + \varepsilon} -
         \left. \frac{\partial\psi}{\partial\phi}
         \right|_{\phi_0 - \varepsilon} \right)
  + \frac{g}{\sqrt{2}\,  \rho}
  \psi(\rho,\phi_0) = 0 \; .
\end{equation*}
\noindent To simplify notation we define $G = \sqrt{2}  g  \rho$
and write $\Delta\Big( \left. \frac{\partial\psi}{\partial\phi}
\right|_{\phi_0} \Big)$ for the $\lim$ construction on the left hand
side. Then for a given value of $g$, the wave function for the
relative motion must fulfill the following condition
\begin{equation}
  \label{eq:condition-G-psi}
  \Delta\left(\left.\frac{\partial\psi}{\partial\phi}\right|_{\phi_0}\right)
  = G\,\psi(\rho,\phi_0) \; .
\end{equation}
This equation specifies the boundary condition on the wave function
that arises from the interaction potential. For all $\phi \not \in
\big\{ \pm \tfrac{\pi}{6}, \pm \tfrac{5\pi}{6} \big\}$ this potential
is zero, and the known $g=0$ wave functions $\psi_\eta(z)\,
\psi_{\nu,\mu}(\rho,\psi)$ that are products of \eqref{eq:cm_wf} and
\eqref{eq:relative_wf} solve the Schr\"odinger equation. Let us see if
the factorized wave function \eqref{eq:relative_wf} is capable of
fulfilling the boundary condition \eqref{eq:condition-G-psi} for
values of $g$ larger than zero. The boundary condition for a wave
function factorized in the radial and angular parts becomes an
equation for the angular function $f(\mu,\phi)$:
\begin{equation}
  \label{eq:condition-G}
  \Delta\left(\left.\frac{\partial f(\mu,\phi)}{\partial\phi}\right|_{\phi_0}\right)
  = G\,f(\mu,\phi_0) \; .
\end{equation}
\noindent The left hand side depends only on $\mu$ and $\phi_0$, but
$G$ also depends on $\rho$, \emph{unless} $g=0$.  This means that
$\rho$ and $\phi$ variables are coupled since $G$ depends on $\rho$,
and that a factorized wave function doesn't generally solve the
problem.

For the sake of argument, let us assume that $G$ is $\rho$-independent
and solve the problem with this assumption. Solving this new (and much
simpler) problem will constitute a `toy model' for the system that
will give us valuable insight into the original problem. The wave
function may now be factorized with the radial part given by the
Laguerre polynomials. The angular part, $f(\mu,\phi)$, bears the
requirements on the wave function from the Pauli principle and parity
such that
\begin{align}
  \label{eq:condition-pauli}
  &f(\mu,\pm \tfrac{\pi}{2}) = 0 && \text{(Pauli principle)} \; ,\\
  \label{eq:condition-parity}
  &f(\mu,-\phi) = \mp f(\mu,\phi) && \text{(even/odd parity)} \; .
\end{align}
Having these symmetries in mind, it suffices to find the angular part
of the wave function only on the first and second domains from
figure~\ref{fig:hypersfaerisk}. On these domains the most general form
of the angular part is
\begin{align*}
  &f(\mu,\phi) = A \, \cos(\mu\phi) + B \, \sin(\mu\phi) \; ,
  &&\phi \in \big[\tfrac{\pi}{6},\tfrac{\pi}{2}\big] \; ,\\
  &f(\mu,\phi) = C \, \cos(\mu\phi) + D \, \sin(\mu\phi) \; ,
  &&\phi \in \big[-\tfrac{\pi}{6},\tfrac{\pi}{6}\big] \; .
  \end{align*}
It can be shown (see appendix \ref{sec:solvedmu}) that a parity state
can fulfill the boundary condition at $\phi=\tfrac{\pi}{6}$ only if
the following equations are satisfied for a given value of $G$
\begin{align}
  \label{eq:determinant-even}
  &\mu \, \sin\left(\mu\tfrac{\pi}{2}\right)
  + G \, \sin\left(\mu\tfrac{\pi}{3}\right)
  \sin\left(\mu\tfrac{\pi}{6}\right) = 0
  &&\text{(even parity)} \\[0.5em]
  \label{eq:determinant-odd}
  &\mu \, \cos\left(\mu\tfrac{\pi}{2}\right)
  + G \, \sin\left(\mu\tfrac{\pi}{3}\right)
  \cos\left(\mu\tfrac{\pi}{6}\right) = 0
  &&\text{(odd parity)}
\end{align}
\begin{figure}[t]
  \centering
  \includegraphics[width=\columnwidth]{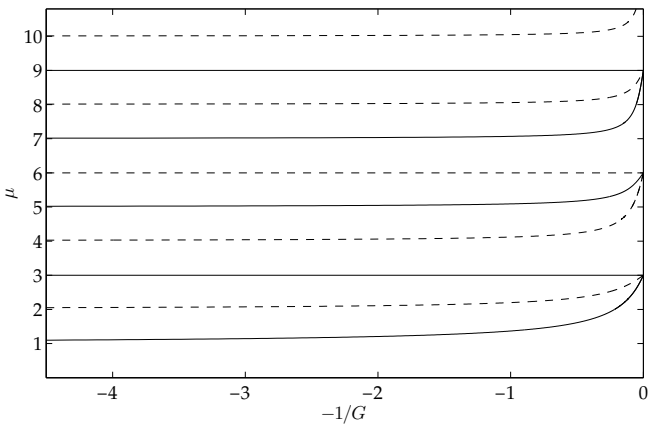}
  \caption{Solutions to \eqref{eq:determinant-even} (even parity,
    dashed lines) and \eqref{eq:determinant-odd} (odd parity, solid
    lines) as functions of $-1/G$.}
  \label{fig:solvedmu_plot}
\end{figure}
\noindent Allowing $\mu \geq 1$ to take non-integer values, the
solutions for a repulsive interaction $G \geq 0$ are seen on
figure~\ref{fig:solvedmu_plot}. This figure is interpreted as the
angular energy spectrum (putting $\nu = \eta = 0$ and neglecting the
constant off-set energy) for our na\"{i}ve `toy model' for the system,
where we neglect the coupling between $\phi$ and $\rho$ for
$g>0$. Generally it is \emph{not} the spectrum for the initial
Hamiltonian, but rather some unknown toy model Hamiltonian that allows
factorized wave functions. It is apparent that the factorized wave
function should solve the initial problem with $G$ dependent on $\rho$
in the non-interacting case $G=0 \Leftrightarrow g=0$.

Solving eqs.~\eqref{eq:determinant-even} and \eqref{eq:determinant-odd}
for the allowed values of $\mu$ in the non-interacting limit
yields\footnote{In this notation $a \equiv_b c$ means $a$ is congruent
  to $c$ modulo $b$.}
\begin{align*}
  &g = 0: \quad
  \begin{cases}
    \mu = 2,4,6\ldots &\hspace{1em} \equiv_2 0 \qquad  \text{(even parity)}\\
    \mu = 1,3,5\ldots &\hspace{1em} \equiv_2 1 \qquad \text{(odd parity)}
  \end{cases}\\[0.5em]
\end{align*}
Combined with eq.~\eqref{eq:condition-parity} this implies that
\begin{align*}
  &g = 0: \quad
  \begin{cases}
    f(\mu,\phi) = \sin(\mu\phi) \; , \quad &\mu \equiv_2 0
    \qquad  \text{(even parity)}\\
    f(\mu,\phi) = \cos(\mu\phi) \; , \quad &\mu \equiv_2 1
    \qquad  \text{(odd parity)}
  \end{cases}
\end{align*}
Solving eqs.~\eqref{eq:determinant-even} and \eqref{eq:determinant-odd}
for the interacting case $g\neq 0$ yields the following integer solutions:
\begin{align*}
  &g \neq 0: \quad
  \begin{cases}
    \mu = 6,12,18\ldots &\equiv_6 0 \qquad  \text{(even parity)}\\
    \mu = 3,9,15\ldots  &\equiv_6 3 \qquad  \text{(odd parity)}
  \end{cases}
\end{align*}
Notice that we generally have, that even (odd) parity solutions have $\mu$
even (odd).

Now, let us discuss the property of the $g\neq 0$ solutions just
obtained. Regardless of parity, they are characterized by $\mu\equiv_3
0$ and seen as the horizontal lines in
figure~\ref{fig:solvedmu_plot}. Thus they are independent of $G$ (and
$g$), so these wave functions also solve the initial problem with
$g>0$. The reason for this is, that the wave function is zero on the
lines of interaction $\phi = \pm \tfrac{\pi}{6}$, i.e. whenever two
particles meet. This is a very important observation, because it means
that $\langle V \rangle = 0$ for all values of $g$. One may say, that
these states never ``feel'' the interaction, and thus never need to
adjust to it. For this reason, we call them the non-interacting
states.

We also see that at $1/G=0$ all other states become degenerate with
these non-interacting states and hence demand the wave function to
vanish whenever two particles meet. In fact, it does not surprise us,
that only the wave functions that vanish on the lines of interaction
are acceptable in the strongly interacting limit, since otherwise
$\langle V\rangle$ would diverge as $G \rightarrow \infty$. This 
has been discussed previously in the context of fermionic systems in
Refs.~\cite{volosniev2013,levinsen2014}

To obtain the full set of solutions for the initial problem (with
$G=\sqrt{2}g \rho$) at $1/g=0$ we note that the corresponding wave
function should also be of factorized form. This observation follows
since eq.~\eqref{eq:condition-G-psi} can only be satisfied for
infinite interaction if the wave function vanishes when two particles
meet. The angular and radial parts becomes independent since two
particles meet on a line which is solely determined by $\phi$. The
only wave functions that are factorized in $\rho$ and $\phi$
coordinates in each region of figure~\ref{fig:hypersfaerisk} and
vanish whenever two particles meet are the non-interacting solutions
obtained above. Thus, if we can construct orthogonal wave functions
using these non-interacting states we actually have an analytic
expression for the wave functions in the strongly interacting
limit~\cite{gharashi2013}.  We take the wave functions for the
non-interacting states and multiply them with a number $a_i$ in every
domain I, II and III on figure~\ref{fig:hypersfaerisk}:
\begin{equation}
  \label{eq:domain-coeff}
  \psi_{\eta}(z) \, \psi_{\nu,\mu \equiv_3 0} (\rho,\phi) \cdot
  \begin{cases}
    a_\text{I} \quad &\text{in I}\\
    a_\text{II} \quad &\text{in II}\\
    a_\text{III} \quad &\text{in III}    
  \end{cases}
\end{equation}
The subscript $\mu \equiv_3 0$ indicates that only these values are
acceptable, while $\nu$ and $\eta$ may still be any non-negative
integer. Obviously, there will be the non-interacting state with the
same wave function for infinite repulsion as in the $g=0$ limit, so
its wave function in the strongly interacting limit must have
$(a_\text{I}, a_\text{II}, a_\text{III}) = (1,1,1)$, i.e. we multiply
by one in every domain. We want to create orthogonal wave functions
with definite parity, and (up to some nonphysical phase factors) this
can only be done by choosing the domain coefficients as $(a_\text{I},
a_\text{II}, a_\text{III}) = (1,-2,1)$ and $(a_\text{I}, a_\text{II},
a_\text{III}) =(1,0,-1)$.\footnote{Which state is odd and which is
  even is not decided by the domain coefficients alone, but also from
  the symmetry/antisymmetry of the $g=0$ wave function over the $\phi
  = 0$ line.} This concludes the construction of the wave functions
for the energy and parity eigenbasis in the strongly interacting
limit.

It is clear that the energy in the two interaction limits is given as
$\tfrac{3}{2} + 2\nu + \mu + \eta$ with the discussed restriction $\mu
\equiv_3 0$ in the strongly interacting limit. Thus, the toy model
spectrum depicted on figure~\ref{fig:solvedmu_plot} reduces to the
correct angular excitation spectrum in these limits. This suggests
that the spectrum for the initial problem should look similar to the
toy model spectrum, as we somehow have to connect these limits to
obtain the spectrum for $0 < g < \infty$. However, bear in mind that
the toy model spectrum, being a function of $G = \sqrt{2} g \rho$,
will force us to pick a value of $\rho$ if we want to map it to a
spectrum depending on $g$. Later, we will do this, but first we will
introduce a more sophisticated way of handling the problem for
intermediate values of the interaction strength.


\begin{figure*}[tbp]
  \centering
  \includegraphics[width=2\columnwidth]{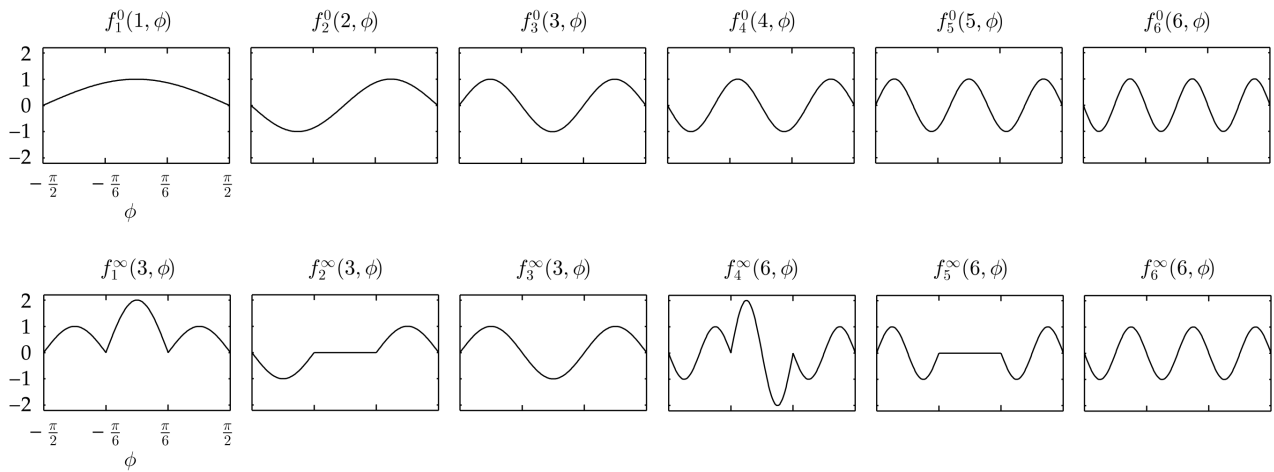}
  \caption{The angular functions for the first six states in the $\mu$
    spectrum in the two limits $g=0$ (upper row) and $1/g=0$.}
  \label{fig:limits_plot}
\end{figure*}

\section{Approximated wave functions}
\label{sec:variation}

In this section we will use the factorized wave function presented in
the previous section to describe the system with $0 \leq g <
\infty$. We will calculate and discuss the energies and probability
densities of the approximated wave functions.

\subsection{Assumptions}

To construct our variational wave function we make two assumptions
about the system: \emph{i)} the first one is  about the
adiabatic connection of the states between $g=0$ and $1/g=0$ limits
where the wave functions should have factorized form as discussed
above, and \emph{ii)} inspired by the discussion in the previous section we
assume that it is more important to describe the angular part of the
wave function, since the interaction happens on $\phi=\phi_0$ line, so
we fix the radial quantum number and find the angular part that
minimizes the energy.

Suppose we have adiabatically evolved the system from the initial
state $\left| \nu,\mu_0, \eta \right>$ at $g=0$ to the state at
$1/g=0$. What would the final state be? Clearly, the center-of-mass
quantum number $\eta$ and the parity eigenvalue would be the same, but
$\nu$ and $\mu$ are generally not good quantum numbers for
intermediate values of $g$ and can in principle be very different in
two limits. However, having in mind the toy model with constant $G$
from the previous section, we assume that the state evolves smoothly
into the wave function which has a spatial profile as similar as
possible to the profile of the initial wave function with changes
happening mostly in the angular part, i.e. we assume that the final
state has $\nu$ unchanged. In the same spirit we assume that $\mu$ is
$\mu_0$ rounded up to the nearest multiple of three. These assumptions
were proven numerically to be true for the lowest part of the energy
spectrum, as we will see later in this report.  To demonstrate this
adiabatic connection we show the angular functions, $f(\mu,\phi)$, for
the first six states in the $\mu$ spectrum in the $g=0$ limit in the
upper row of figure \ref{fig:limits_plot}. In the lower row we have
the first six angular functions at $1/g=0$ constructed from the $g=0$
angular functions by using the domain coefficients discussed
above. The functions are labeled $f_{\mu_0}^g (\mu,\phi)$, where
$\mu_0$ is a label to keep track of the states as there are three
different states for every allowed value of $\mu$ at $1/g=0$, in our
assumption the label is the $g=0$ value for the angular excitation
quantum number.  Note that the parity eigenvalue is given by
$(-1)^{\mu_0}$.

\begin{figure*}
    \centering
    \includegraphics[width=.98\columnwidth]{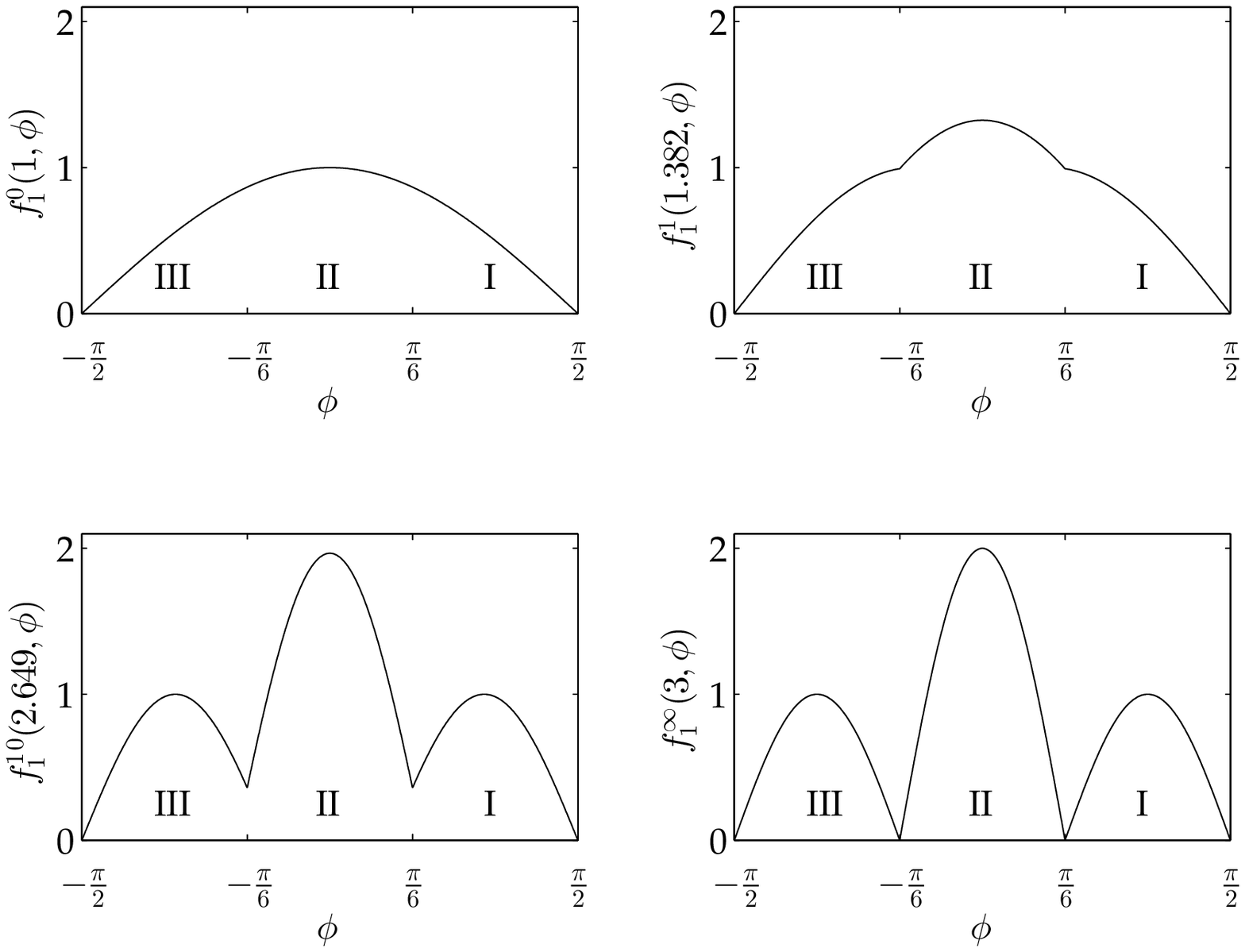}
    \qquad
    \includegraphics[width=.98\columnwidth]{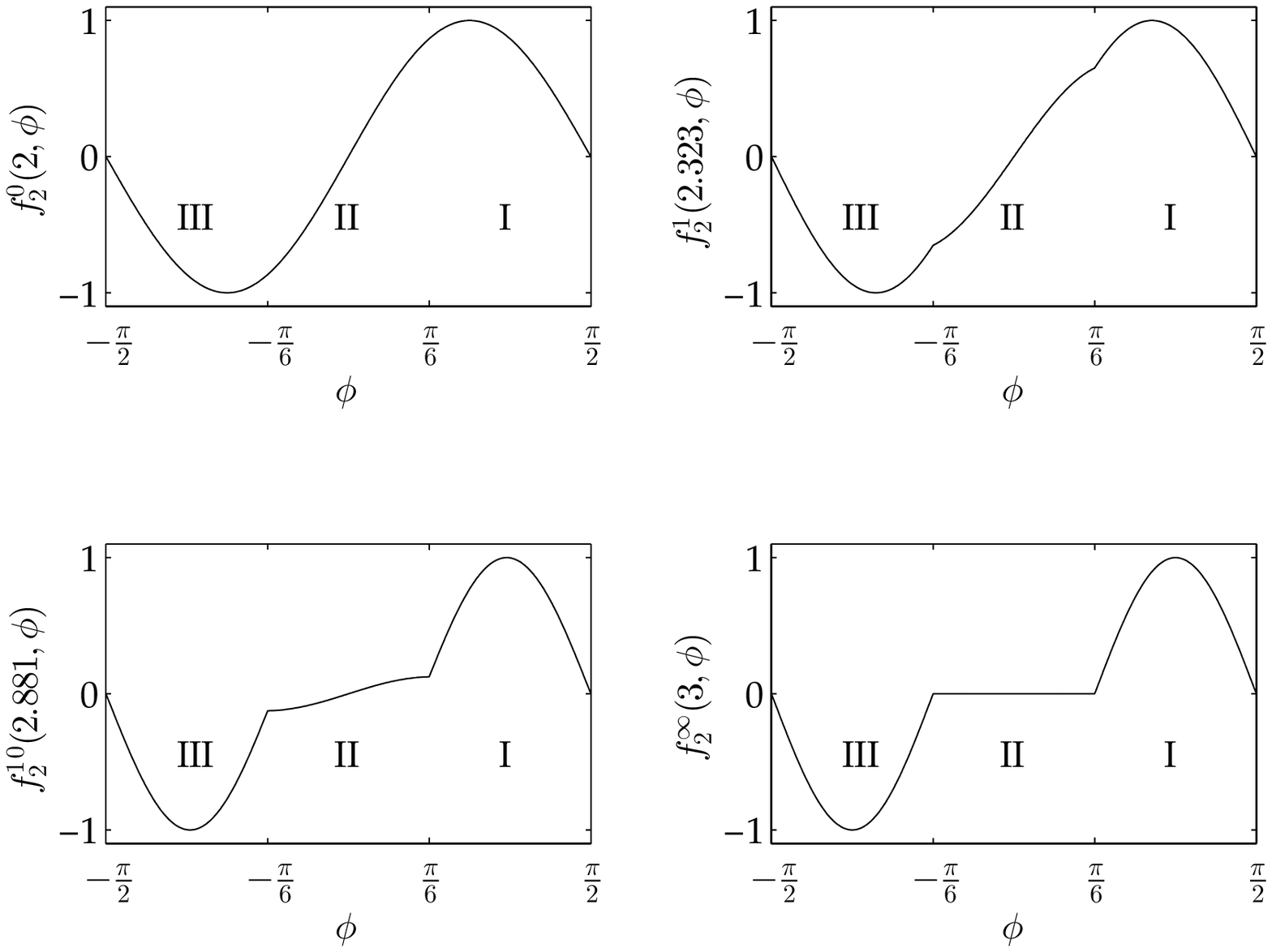}
    \caption{The angular function for the ground state, $f_1^g
      (\mu,\phi)$ (left side, odd parity), and the first excited state,
      $f_2^g (\mu,\phi)$ (right side, even parity), for $g=0,1,10$ and
      $1/g=0$. The corresponding values of $\mu $ are in interval
      $[1,3]$ and $[2,3]$, respectively. It is explained in the text
      how $\mu$ is chosen for every value of $g$.}
    \label{fig:wavefunctions}
\end{figure*}

As we have already noted in the previous section, the wave function
for the relative motion can in general only be of the factorized form
\eqref{eq:relative_wf} when $g=0$ or $1/g=0$. Nevertheless, since we
are only seeking an approximate solution to the interacting state wave
functions, we assume a factorized relative wave function also for
intermediate values of the interaction strength.  Assuming that any
state is characterized by constant $\nu$, $\mu_0$ and $\eta$ we denote
this state by $\left| \nu,\mu_0,\eta \right>_g^{\text{ap}}$, where we
have put an superscript `ap' on the state ket to indicate the
approximation. We now aim to find a reasonable form for the wave
function. If we consider the individual states in
figure \ref{fig:limits_plot}, we can qualitatively understand how the
wave function behaves for $0 < g < \infty$. The wave function is
forced to vanish for $\phi = \pm \tfrac{\pi}{6}$, and so we can
imagine gripping these points on the $g=0$ wave function and slowly
pulling them down towards zero when $g$ increases. Pursuing this idea,
we construct angular functions with this property that reduces to the
angular functions characterized by $\mu_0$ in the limits, i.e.
\begin{align}
  \label{eq:assume_wf}
  \left<\rho,\phi,z \, \left| \right. \! \nu,\mu_0,\eta
  \right>_g^\text{ap}=\psi_\eta(z) R_\nu (\mu,\rho) f_{\mu_0}^g (\mu,\phi)
\end{align}
where the radial part given as in eq.~\eqref{eq:relative_wf}
\begin{align}
  R_\nu (\mu,\rho) =
  A\, L_\nu^{(\mu)} \! \big(\rho^2\big) \, e^{-\rho^2 / 2} \, \rho^\mu \,  \; .
\end{align}
The true dependency on $(\rho,\phi)$ is unknown for any $g$ from the
interval $]0,\infty[$, but since the above functional form is correct
at the boundaries of this interval, we take it as a reasonable
approximation. We assume that $\mu$ is a continuous variable and a
function of the interaction strength, $\mu = \mu(g)$. It looses its
meaning as a quantum number for the intermediate values of $g$, and we
think of it as a variational parameter. For a given value of $g$ we
will vary $\mu(g)$ such that the energy matrix element with the
approximated wave function is minimal. We again would like to stress
the difference between the state labeling number $\mu_0$ and the
variational parameter $\mu(g)$. For the ground state ($\mu_0 = 1$) the
limits are $\mu(0) = 1$ and $\mu(\infty) = 3$, and for the first
excited state ($\mu_0 = 2$), we have $\mu(0) = 2$ and $\mu(\infty) =
3$. In general, a state labeled by $\mu_0 = 1, 2, 3, \dots$ has by
definition $\mu(0) = \mu_0$, and $\mu(\infty)$ is $\mu_0$ rounded up
to the nearest multiple of three.  With these assumptions, it can be
shown (see appendix~\ref{sec:angular}) that
\begin{align}
  \label{eq:angular}
  & f_{\mu_0}^g (\mu,\phi) \\ &=
   \begin{cases}
     \sin\left(\mu\left(\tfrac{\pi}{2}-\phi\right)\right)
     &\text{in I}\\[0.5em]
     (-1)^{\mu_0+1} \sin\left(\mu\left(\tfrac{\pi}{6}-\phi\right)\right)
   + \sin\left(\mu\left(\tfrac{\pi}{6}+\phi\right)\right)
     &\text{in II}\\[0.5em]
     (-1)^{\mu_0+1}
     \sin\left(\mu\left(\tfrac{\pi}{2}+\phi\right)\right)
     &\text{in III}
   \end{cases} \nonumber
\end{align}
is an angular function with the right parity that reduces to the
solutions in the limits $g \rightarrow 0$ and $g \rightarrow \infty$
that we assume are adiabatically connected. It has been found by
proposing a general ansatz function with the desired parity and the
property that it vanishes at $\phi = \pm \tfrac{\pi}{2}$ as required
by the Pauli principle, and at $\phi = \pm \tfrac{\pi}{6}$ for $\mu =
3,6,9,\dots$ For a few values of $\mu$, this function is plotted in
figure~\ref{fig:wavefunctions}, where we clearly see the discussed
behavior.

With the assertion of \eqref{eq:assume_wf} as the wave function for
the relative motion, the approximation for the full wave function
including the center-of-mass part becomes
\begin{align}
  \label{eq:approx_wf}
  \psi(\rho,\phi,z)
  \approx
  \psi_\eta(z) \, R_\nu (\mu,\rho) \, f_{\mu_0}^g ( \mu,\phi) \; ,
\end{align}

\subsection{The Hamiltonian matrix elements}

The approximated interacting state wave functions \eqref{eq:approx_wf}
and the non-interacting state wave functions form a basis in which we
would like to investigate the representation of the Hamiltonian. Thus
we calculate the matrix elements for $H$ in this basis. This serves
two purposes: \emph{i)} we need the expectation value of a given approximated
wave function to be able to variationally determine the correspondence
between $g$ and $\mu$, and \emph{ii)} we want to diagonalize the Hamiltonian
in a selected subset of basis wave functions, thus getting even closer
to the correct wave functions and the energy spectrum. We consider
three cases for the states involved in the matrix element. Firstly, if
the two states are both non-interacting, we obviously get
\begin{align}
  \label{eq:matrixel-nonnon}
  &\left< \nu',\mu_0',\eta' \vphantom{H} \right|
  H \left| \nu,\mu_0,\eta \vphantom{H} \right> \nonumber\\[0.3em]
  &= \left(\tfrac{3}{2} + 2\nu + \mu_0 + \eta\right)
  \delta_{\nu'\nu}\,\delta_{\mu_0'\mu_0}\,\delta_{\eta'\eta}
\end{align}
as they are both eigenstates of $H$ for all values of $g$. Secondly,
the matrix element between an interacting and a non-interacting state
is zero. This is shown in Appendix~\ref{sec:energy}, where we also
show that the matrix element between two approximated interacting
states is given as

\begin{widetext}
  \begin{align}
    \label{eq:matrixel-intint}
    \tensor*[^{\text{ap}}_g]
    {\left<\nu',\mu_0',\eta' \vphantom{H\mu_0'} \right| H \left|
        \vphantom{H\mu_0'} \nu,\mu_0,\eta \right>}
    {^{\text{ap}}_g}
    &=
    \tensor*[^{\text{ap}}_g]
    {\left<\nu',\mu_0',\eta' \vphantom{H_0\mu_0'} \right| H_0 \left|
        \vphantom{H_0\mu_0'} \nu,\mu_0,\eta \right>}
    {^{\text{ap}}_g}
    +
    \tensor*[^{\text{ap}}_g]
    {\left<\nu',\mu_0',\eta' \vphantom{V\mu_0'} \right| V \left|
        \vphantom{V\mu_0'} \nu,\mu_0,\eta \right>}
    {^{\text{ap}}_g}
    \qquad \text{with} 
    \\[1em]
    \label{eq:H0-intint}
    \tensor*[^{\text{ap}}_g]
    {\left<\nu',\mu_0',\eta' \vphantom{H_0\mu_0'} \right| H_0 \left|
        \vphantom{H_0\mu_0'} \nu,\mu_0,\eta \right>}
    {^{\text{ap}}_g}
    &= \delta_{\eta'\eta}
    \left(\tfrac{3}{2} + 2\nu + \mu + \eta \right)
    \int\limits_0^\infty \mathrm{d} \rho\,\rho\, R_{\nu'}(\mu',\rho) R_\nu(\mu,\rho)
    \int\limits_{-\pi}^{\pi} \mathrm{d} \phi \, f_{\mu_0'}^g (\mu',\phi)
    f_{\mu_0}^g (\mu,\phi) \nonumber\\
    & + \delta_{\eta'\eta}
    \left(1+(-1)^{\mu_0'+\mu_0}\right) \mu \sin\left(\mu'\tfrac{\pi}{3}\right)
    \left(2\cos\left(\mu\tfrac{\pi}{3}\right) + (-1)^{\mu_0}\right)
    \int\limits_0^\infty \mathrm{d} \rho\,\rho^{-1}\, R_{\nu'}(\mu',\rho) R_\nu(\mu,\rho)
    \; , \\[1em]
    \label{eq:V-intint}
    \tensor*[^{\text{ap}}_g]
    {\left<\nu',\mu_0',\eta' \vphantom{V\mu_0'} \right| V \left|
        \vphantom{V\mu_0'} \nu,\mu_0,\eta \right>}
    {^{\text{ap}}_g}
    &= \delta_{\eta'\eta} \,
    g \, \sqrt{2}
    \left(1+(-1)^{\mu_0'+\mu_0}\right) \sin\left(\mu'\tfrac{\pi}{3}\right) \,
    \sin\left(\mu\tfrac{\pi}{3}\right)
    \int\limits_0^\infty \mathrm{d} \rho\, R_{\nu'}(\mu',\rho) R_\nu(\mu,\rho)
    \; .
  \end{align}
\end{widetext}

\noindent Note that the matrix elements between two states with
different center-of-mass excitation or different parity always
vanishes, as it should.  Also note that the states with different
values of $\nu$ can mix with one another.

For a given interacting state and a given value of $g$, we need a
criteria to choose the value of $\mu$. It seems natural to take a
variational approach to this problem. For a given value of the
interaction strength $g$, we will consider the diagonal matrix element
for that state, i.e. its expectation value, $ \epsilon \equiv \langle
H \rangle = \tensor*[^{\text{ap}}_g] {\left<\nu,\mu_0,\eta
    \vphantom{H\mu_0} \right| H \left| \vphantom{H\mu_0}
    \nu,\mu_0,\eta \right>}{^{\text{ap}}_g}$, and vary $\mu$ such that
this expectation value is minimal. Since $\epsilon$ is linear in $g$,
but a very complicated function of $\mu$, we pull $g$ outside the
interaction term such that $\langle V \rangle = g \langle V'
\rangle$. Notice that $\langle H_0 \rangle$ and $\langle V' \rangle$
are determined solely by $\mu$. If for all $g \in ]0,\infty[$ there is
a local minimum in the trial energy $\epsilon (\mu)$ at some value
$\mu_\text{min}$, we can find $\mu_\text{min}$ for every value of $g$ from equation
\begin{align*}
    \frac{\mathrm{d}\epsilon}{\mathrm{d}\mu}
    \! \left. \vphantom{\frac{\mathrm{d}}{\mathrm{d}\mu}}\right|_{\mu_\text{min}}
  = \left.\frac{\mathrm{d}}{\mathrm{d}\mu} \left< H_0 \right>\right|_{\mu_\text{min}}
  + g \left.\frac{\mathrm{d}}{\mathrm{d}\mu} \left< V' \right>\right|_{\mu_\text{min}} = 0.
\end{align*}
\noindent This gives us the relationship between $\mu$ and $g$ that
minimizes $\epsilon$:
\begin{align}
  \label{eq:minimizing-g}
  g(\mu) =
  - \, \frac{ \frac{\mathrm{d}}{\mathrm{d}\mu} \left< H_0 \right> }
  { \frac{\mathrm{d}}{\mathrm{d}\mu} \left< V'\right> }  \; .
\end{align}
Using \eqref{eq:H0-intint} and \eqref{eq:V-intint} it is possible to
find an analytic form for this expression.

For the ground state $\left|0,1,0\right>_g^\text{ap}$ and first
excited state in the $\eta = 0$ spectrum,
$\left|0,2,0\right>_g^\text{ap}$, we use eqs.~\eqref{eq:H0-intint}
and~\eqref{eq:V-intint} to calculate analytic expressions for $\langle
H_0 \rangle$ and $\langle V \rangle$, as shown in
Appendix~\ref{sec:energy}. Then eq.~\eqref{eq:minimizing-g} is used to
establish the energy minimizing relation between $g$ and $\mu$. The
analytic expression for $g(\mu)$ is very lengthy and not very
informative, so we will not quote it here. The equation gives the
value of $g$ for a given $\mu_\text{min}$, but one would rather
provide a value for the interaction strength $g$ and find
$\mu_\text{min}$. This is achieved simply by numerically finding the
root in $g(\mu)~-~g$. For $\left|0,1,0\right>_g^\text{ap}$ and
$\left|0,2,0\right>_g^\text{ap}$ we use this method to compute a list
of $\mu$'s for different values of $g$. Writing $\mu = \mu_0 + \delta
\mu$, we show a set of $\delta \mu$'s for some values of $g$ in Table
\ref{tab:deltamu}, some of which were used when plotting the angular
functions in figure \ref{fig:wavefunctions}. We could do this for
other interacting states too, but for the sake of argument we will use
$\delta \mu$'s minimizing the energy of
$\left|0,1,0\right>_g^\text{ap}$ for all states with $\mu_0 \equiv_3
1$ and the $\delta \mu$'s for $\left|0,2,0\right>_g^\text{ap}$ for all
states with $\mu_0 \equiv_3 2$. This approach yields very accurate
results for the energy and the wave function, so
Table~\ref{tab:deltamu} provide a very simple access to accurate
three-body wave functions without any further calculations.
\begin{table}[htbp]
  \centering
  \begin{tabular}{lll}
    \toprule
        \quad & $\mu_0 \equiv_3 1$
        \quad & $\mu_0 \equiv_3 2$ \\[0.3em]
     $g$ & $\delta \mu$ & $\delta \mu$\\
    \midrule
    $0$      & $0$       & $0$ \\
    $1/4$    & $0.08759$ & $0.09054$ \\
    $1/3$    & $0.11835$ & $0.11957$ \\
    $1/2$    & $0.18186$ & $0.17547$ \\
    $1$      & $0.38192$ & $0.32278$ \\
    $2$      & $0.75567$ & $0.52664$ \\
    $10/3$   & $1.09918$ & $0.67518$ \\
    $10$     & $1.64939$ & $0.88054$ \\
    $20$     & $1.82016$ & $0.93948$ \\
    $100$    & $1.96352$ & $0.98782$ \\
    $10000$  & $1.99963$ & $0.99988$ \\
    $\infty$ & $2$       & $1$\\
    \bottomrule
  \end{tabular}
  \caption{Energy minimizing values of $\delta \mu$ for
    $\left|0,1,0\right>_g^\text{ap}$ (second column) and
    $\left|0,2,0\right>_g^\text{ap}$ (third column). The numbers are
    used for all states with the indicated $\mu_0$.}
  \label{tab:deltamu}
\end{table}

\begin{figure*}[htbp]
  \centering
 \includegraphics[width=2\columnwidth]{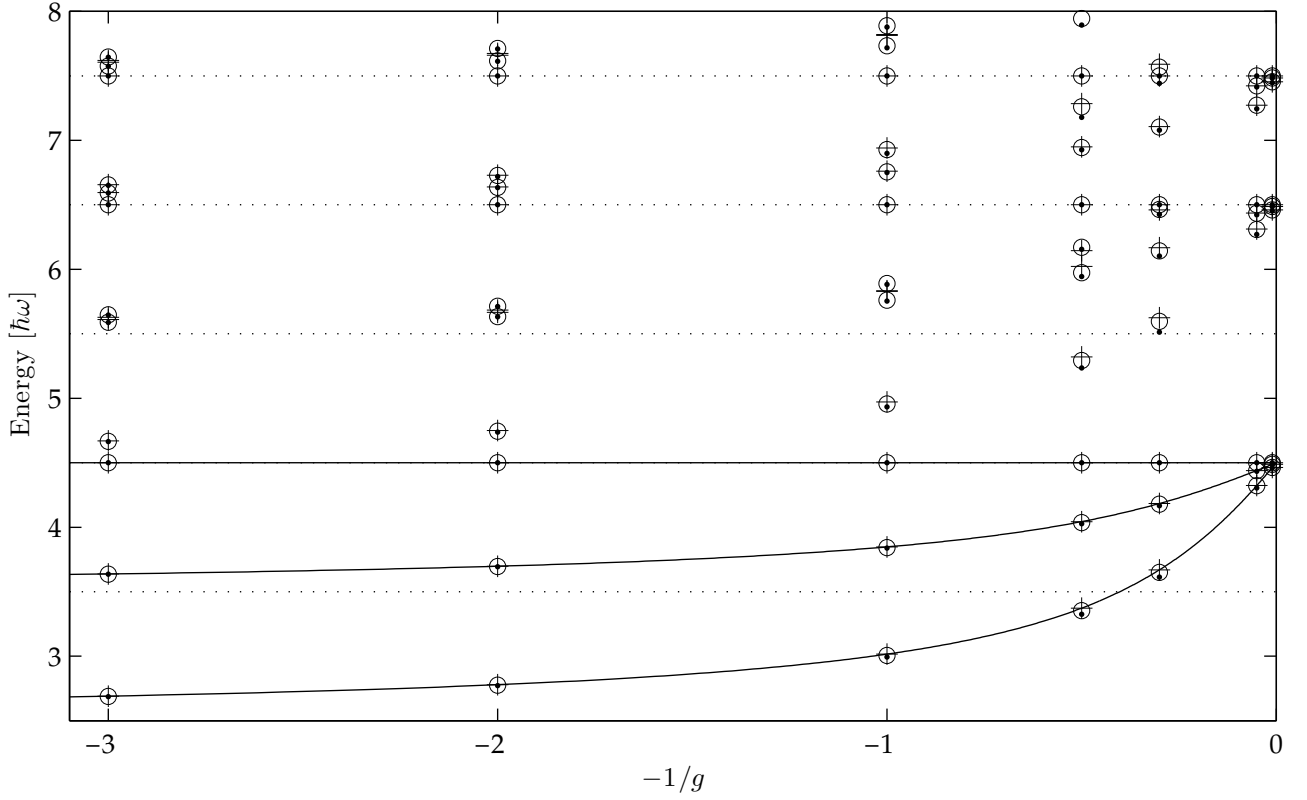}
 \caption{Diagonal elements, i.e. the expectation values $\epsilon$,
   ($+$ and lines) and eigenvalues ($\circ$) of the Hamiltonian in a
   basis of 54 approximated eigenstates compared to the exact energies
   ($\cdot$). We only show the lowest part of the spectrum with $\eta
   = 0$. We see that the deviation of the expectation values and
   eigenvalues from the exact energies is largest at $1/g=0.5$ and
   $1/g=0.3$.}
  \label{fig:energy}
\end{figure*}

\subsection{Diagonalization of the Hamiltonian}
Using the factorized function \eqref{eq:approx_wf} with the angular
parameters, $\mu$, taken from Table~\ref{tab:deltamu}, we are able to
create an arbitrarily large basis of functions in which we can
diagonalize the Hamiltonian for different values of $g$. As an
example, we take a basis of 54 states consisting of
$\left|\nu,\mu_0,\eta\right>_g^\text{ap}$ with $\nu \in \{0,\dots,5
\}$, $\mu_0 \in \{1,\dots,9 \}$ and since states with different
center-of-mass excitation do not couple, we take $\eta = 0$ for all
the states. The resulting energy spectrum is shown in figure
\ref{fig:energy}, where we show the eigenvalues from the
diagonalization ($\circ$) together with the diagonal elements ($+$)
discussed in the previous subsection, versus $-1/g$. Since we
calculated analytic expressions for $\epsilon(\mu)$ for the ground and
first excited state (see appendix~\ref{sec:energy}), we show the
spectrum for these states, including the constant energy for the
second excited state, as solid lines.  The figure also contains the
exact energies ($\cdot$) calculated numerically by diagonalizing the
total Hamiltonian for different values of $g$ in a large basis of
eigenstates for $ H_0$ \cite{lindgren2014}. This is a fairly large
computational task, and so we are interested in seeing how well the
expectation values of the approximated wave functions presented here
compare to the correct energies. It is seen that the expectation
values, $\epsilon$, for the approximated wave functions are nearly
spot-on the correct energies, and that the result after
diagonalization with 54 states is even better, as expected. The
expectation values are a little higher than the true energies, which
is consistent with the presented variational approach.

Notice that when $g \sim 1$ then $\rho$ is dominating the value of $G
= \sqrt{2} \, g \, \rho$, but when $g$ is either very small or very
large, the value of $\rho$ becomes less important and the true wave
function can be very well reproduced using the presented trial wave
function. Thus when $g \sim 1$ the factorized function
\eqref{eq:approx_wf} is further away from the true wave function, and
we expect the deviation of the expectation values $\epsilon$ to be
largest in this region. However, from figure~\ref{fig:energy} we see
that the largest deviation is in fact found for the two sets of data
points at $1/g = 0.5$ and $1/g = 0.3$.

The introduced basis of approximated wave functions has several
advantages, most notable, of course, that it reduces to the eigenbasis
of the Hamiltonian in the limits of weak and strong repulsion, and
hence that $H$ becomes `more and more' diagonal as we approach these
limits. But we also took great advantage of the separability of the
center-of-mass term and the parity symmetry of the Hamiltonian, which
significantly reduced the number of non-zero off-diagonal elements. We
also observe that the factorized functions \eqref{eq:approx_wf} describe
the exact wave functions very accurately which has the consequence
that these factorized functions are coupled weakly by the Hamiltonian,
or in other words: an eigenvector expanded in the approximated states
consists nearly solely of one state.

\begin{figure}[htbp]
  \centering
  \includegraphics[width=\columnwidth]{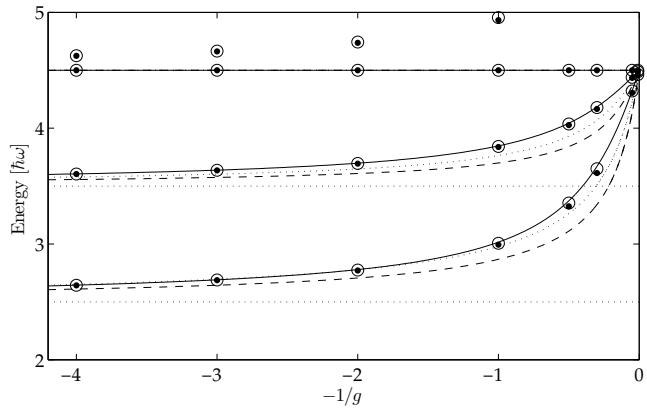}
  \caption{Comparison between the toy model spectrum found from the
    solutions to \eqref{eq:determinant-even} and
    \eqref{eq:determinant-odd} with $G = \sqrt{2} g$ (dotted lines),
    $G = g$ (dashed lines), diagonal elements (solid lines) and
    eigenvalues ($\circ$) of the Hamiltonian as in
    figure~\ref{fig:energy} and the exact energies ($\cdot$).}
  \label{fig:toymodel_comparison}
\end{figure}

For the sake of completeness, we also want to compare the `toy model'
energy spectrum from figure~\ref{fig:solvedmu_plot} with the exact and
the approximated energies from figure~\ref{fig:energy}. However, to do
the comparison, we need to relate $g$ and $G = \sqrt{2} g \rho$ by
choosing some value of $\rho$, see
figure~\ref{fig:toymodel_comparison} for the choices $G = \sqrt{2} g$
(dotted lines) and $G = g$ (dashed lines). We compare the low energy
part of these spectra with the exact energies ($\cdot$), the
expectation values for the approximated wave functions (solid lines)
and the eigenvalues found from diagonalization ($\circ$), just as in
figure~\ref{fig:energy}. The two fixed values of $\rho$ where chosen
arbitrarily, other choices yield similar spectra, and we see no
particular reason for why one choice produces a more accurate
spectrum than another. Surely, we retrieve the correct energies in the
interaction limits, but we also find, that the toy model is quite accurate
for intermediate values of $g$ for these particular choices of
$\rho$. However, different choices of $\rho$ would `stretch' the toy
model spectrum, but not change its general form. This means that even
though the very simple approach to the problem of a general
interaction is not a scheme for finding the correct numerical values
for the energies, we may certainly learn a lot about the shape of the
spectrum and thus the behavior of the system.

\begin{figure*}[htbp]
  \centering
  \includegraphics[width=.98\columnwidth]{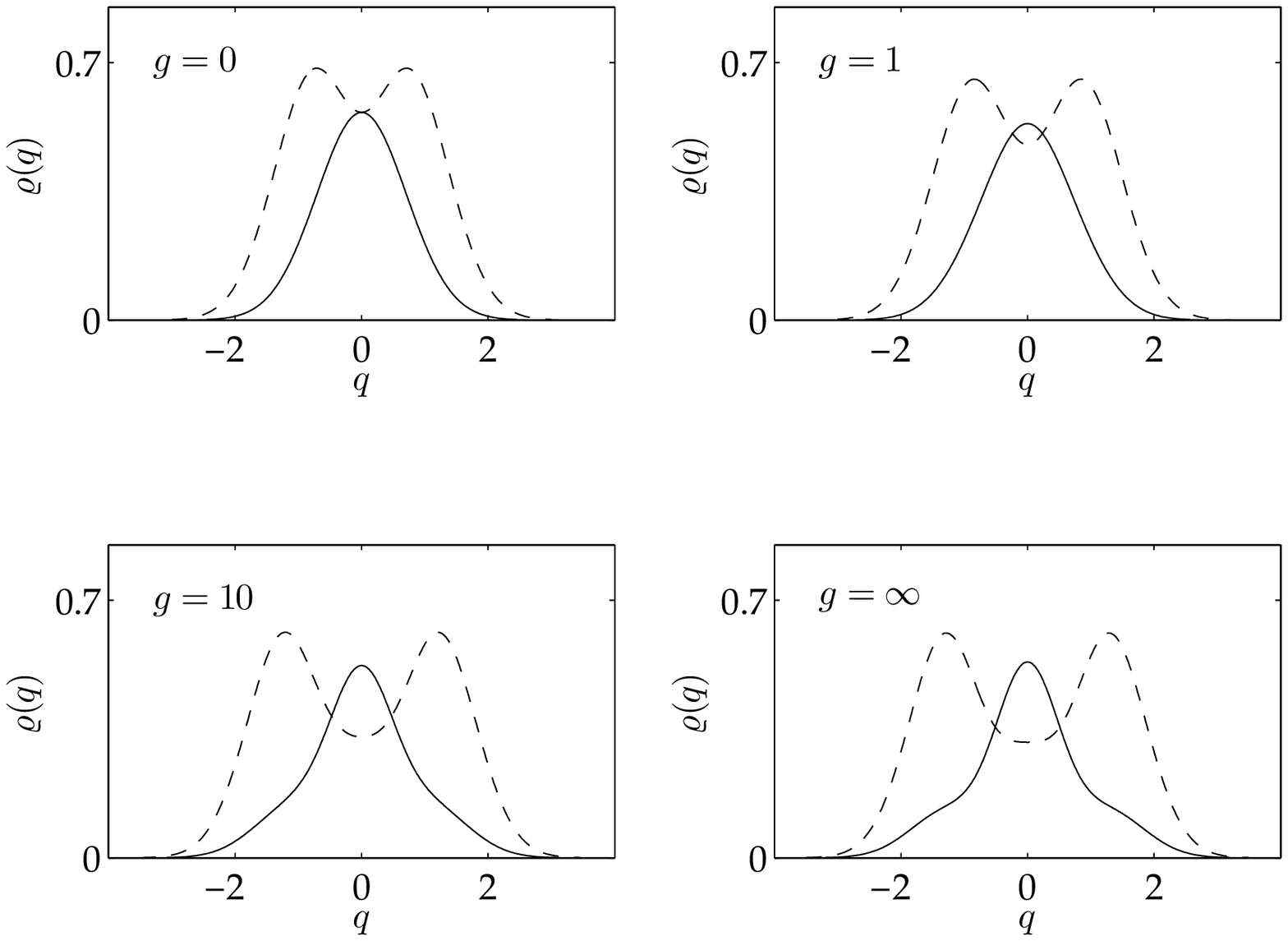}
  \qquad
  \includegraphics[width=.98\columnwidth]{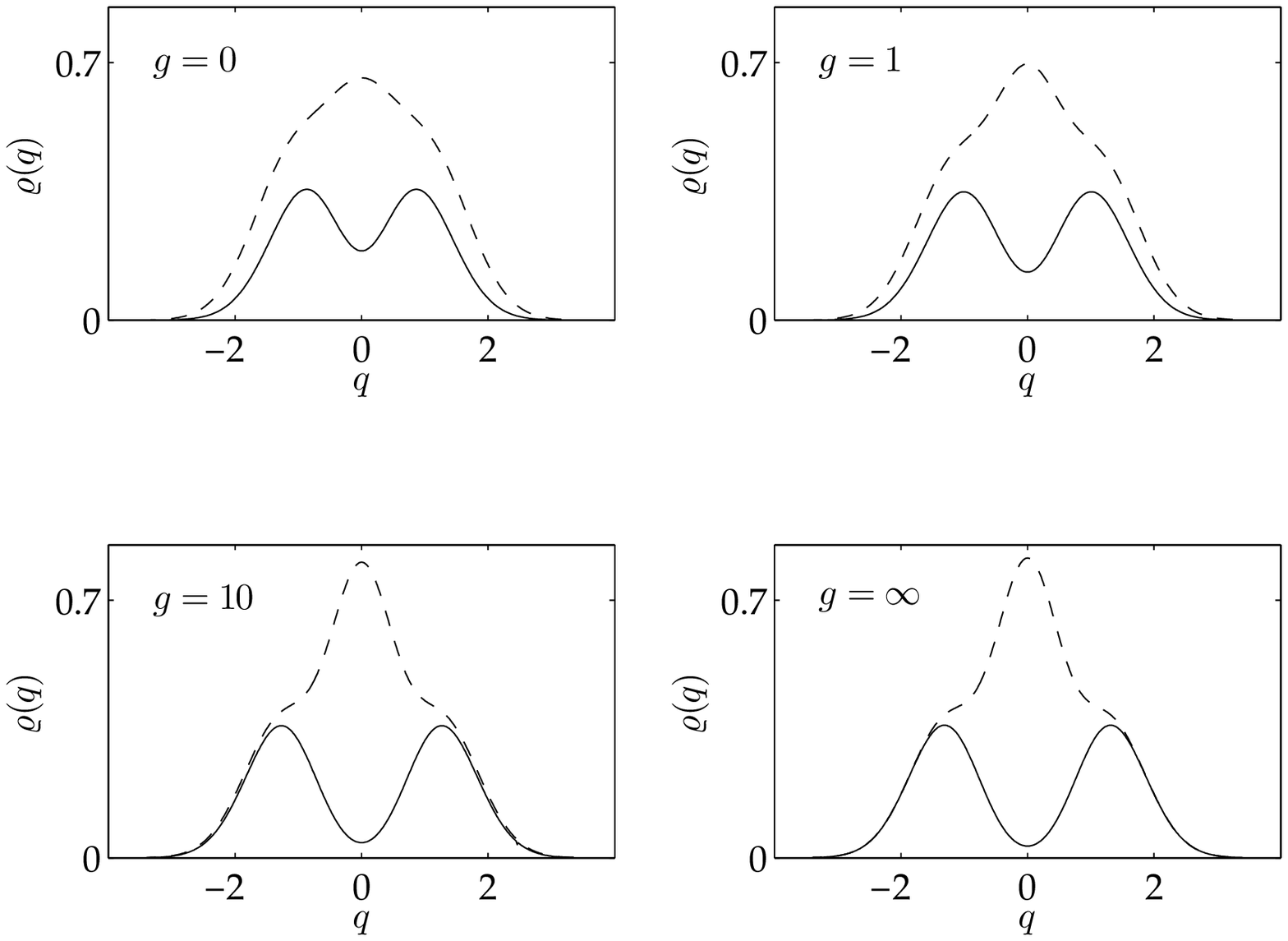}
  \caption{Probability density for particle 3 (solid) and particle 1
    and 2 (dashed) for the ground state
    $\left|0,1,0\right>_g^\text{ap}$ (left side) and for the first
    excited state $\left|0,2,0\right>_g^\text{ap}$ (right side).}
  \label{fig:density}
\end{figure*}

\subsection{Probability densities}
\label{sec:density}

We would like to calculate the probability density of particle $i$ as
a function of $q_i$ using the approximated wave functions for the
ground state $\left|0,1,0\right>_g^\text{ap}$ and first excited state
$\left|0,2,0\right>_g^\text{ap}$. Using the inverse coordinate
transformations, we can write the wave functions in variables
$(q_1,q_2,q_3)$. The desired probability density can be calculated as
\begin{align*}
  \varrho_i(q_i) = \int\limits_{-\infty}^\infty \mathrm{d} q_j
  \int\limits_{-\infty}^\infty \mathrm{d} q_k
  \left|\left<q_1,q_2,q_3 \, \left|
      \right. \! 0,\mu_0,0 \right>_g^\text{ap}\right|^2 \; ,
\end{align*}
where $i,j,k \in \{1,2,3\}$ are all different. Since the function
under the integral is piecewise-defined on domains I, \dots, VI,
we must write the integral as a sum with one term for each domain. The
domains are easily parametrized in the $(q_1,q_2,q_3)$ coordinates,
and for the probability density of particle 3 we get:
\begin{align*}
  \varrho_3(q_3)
  &= 2 \int\limits_{-\infty}^{q_3} \mathrm{d} q_1
       \int\limits_{q_1}^{q_3}    \mathrm{d} q_2
       \left|\psi_\text{I}\right|^2
  + 2 \int\limits_{-\infty}^{q_3} \mathrm{d} q_1
      \int\limits_{q_3}^{\infty}  \mathrm{d} q_2
     \left|\psi_\text{II}\right|^2  \\
  &+ 2 \int\limits_{q_3}^{\infty} \mathrm{d} q_1
       \int\limits_{q_1}^{\infty} \mathrm{d} q_2
      \left|\psi_\text{III}\right|^2 \; ,
\end{align*}
where we have written $\psi_\text{I}$ for the wave function in domain
I, and so forth, note that the contribution from the other domains,
i.e. IV, V, VI is obtained from the invariance of the integrand under
$x \mapsto -x$. When $q_2 \geq q_1$ we integrate over the $x \geq 0$
half plane to find the probability density of particle 1:
\begin{align*}
  \varrho_1(q_1)
  &= \int\limits_{q_1}^{\infty} \mathrm{d} q_2
     \int\limits_{q_2}^{\infty} \mathrm{d} q_3
    \left|\psi_\text{I}\right|^2
  + \int\limits_{q_1}^{\infty} \mathrm{d} q_2
    \int\limits_{q_1}^{q_2}   \mathrm{d} q_3
    \left|\psi_\text{II}\right|^2  \\
  &+ \int\limits_{q_1}^{\infty}  \mathrm{d} q_2
     \int\limits_{-\infty}^{q_1} \mathrm{d} q_3
    \left|\psi_\text{III}\right|^2  \; .
\end{align*}
When $q_2 \leq q_1$ we must integrate over the $x \leq 0$ half
plane. However, we can also use that the probability density must be
invariant under reflection of the $q$ axis in which case we cover the
situation $q_2 \leq q_1$ with $\varrho_1(-q_1)$.  Notice that the
probability density of particle 2 is equal to that of particle 1,
i.e. $\varrho_2(q_2) = \varrho_1(q_1)$, since two particles are
identical. The probability densities are calculated numerically and
normalized such that the integral over all densities is the number of
particles. They are plotted for the ground state and for the first
excited state in figure~\ref{fig:density}.

If the particles are spin-$\tfrac{1}{2}$ fermions, the probability
densities yield some interesting magnetic behavior of the states. Say
that particles 1 and 2 are indistinguishable because they are in the
same spin state, for instance spin-up indicated on
figure~\ref{fig:hypersfaerisk}, opposite to the spin-down state of
particle 3.  For the ground state shown on the left half of
figure~\ref{fig:density} we see that when $g$ increases the two
indistinguishable particles are ``pushed'' to either side of particle
3 which leads to an increasing probability to find the system in the
configuration $\uparrow\downarrow\uparrow$. Of course, particles 1 and
2 cannot be pushed completely away from particle 3 because the energy
contribution from the harmonic trap potential at some point becomes
too great. But as the interaction strength $g$ increases, particle 1
and 2 are pushed farther out to the sides of the trap. This can be
interpreted as ``antiferromagnetic'' behavior of the few-body system,
since the most energy favorable configuration is the one with
alternating spin orientations along the $q$-axis. For the first
excited state shown on the right half of figure~\ref{fig:density}, the
situation is completely different as the configuration
$\uparrow\downarrow\uparrow$ becomes less probable when $g$
increases. In fact, from earlier we know that in the strongly
interacting limit $g \to \infty$, the probability completely vanishes
for this configuration. However, particle 3 may still be found in the
middle of the harmonic trap, though the probability for doing so is
very small as seen on the figure. On the other hand, the probability
for finding particle 1 or 2 here is strongly favored.  Particle 3 is
pushed to one of the sides of the trap and the particles with the same
spin orientation are located next to each other, so the first excited
state exhibits ``ferromagnetic'' behavior. It is quite exciting to see
that a small system of only three particles exhibit increasing
magnetic behavior when the interaction increases. Strongly interacting
particles play an important role in theories of magnetism in condensed
matter physics, but the theories often consider the average behavior
of many particles interacting with each other. Thus, the study of
small interacting system, that are solvable in the interaction limits,
like the system treated here, might lead to a better understanding of
solid-state phenomena.

\section{Mass imbalance}
\label{sec:masses}

Let us returning to the general three-body Hamiltonian consisting of
the terms \eqref{eq:2} and \eqref{eq:3}.  Contrary to the preceding
sections, the masses $m_i$ for $i\in \{1,2,3\}$ are now allowed to be
different. Again, we define a set of dimensionless coordinates
$\bf{q}$ and $\bf{p}$:
\begin{equation*}
  q_i = \frac{\tilde q_i}{\sigma}
  \quad \text{and} \quad
  p_i = \frac{\tilde p_i \, \sigma}{\hbar} \; .
\end{equation*}
where $\sigma=\sqrt{\hbar/\mu_{123}\omega}$ is a length scale and
$\mu_{123}=\sqrt{m_1m_2m_3/(m_1+m_2+m_3)}$. Notice that the 
definition of $\sigma$ is now different compared to the one in Section
\ref{sec:system}, since here we use a sort of reduced mass $\mu_{123}$
instead of the common mass $m$. Again we choose units such that $\hbar
\equiv \omega \equiv 1$, but for pedagogical reasons we do not set
$\mu_{123}$ to unity, and so contrary to before, we cannot set
$\sigma$ to unity. Like in Section \ref{sec:system}, we rotate the
coordinates and get a new set of coordinates $\mathbf{r}= \mathbf{J}
\mathbf{q}$, this time the transformation is given by
\begin{equation*}
  \begin{bmatrix} x \\[0.5em]
                  y \\[0.5em]
                  z \end{bmatrix} = \frac{1}{\sqrt{\mu_{123}}}
\begin{bmatrix}
\mu_{12} & -\mu_{12} & 0 \\[0.5em]
\frac{\mu_{123} m_1}{\mu_{12}M_{12}} & \frac{\mu_{123} m_2}{\mu_{12}M_{12}} & -\frac{\mu_{123}}{\mu_{12}} \\[0.5em]
\frac{m_1}{\sqrt{M_{123}}} & \frac{m_2}{\sqrt{M_{123}}} & \frac{m_3}{\sqrt{M_{123}}}
\end{bmatrix}
\begin{bmatrix} q_1 \\[0.5em]
                    q_2 \\[0.5em]
                    q_3 \end{bmatrix} \; ,
\end{equation*}
where $M_{12}=m_1+m_2$, $M_{123}=m_1+m_2+m_3$ and
$\mu_{12}=\sqrt{m_1m_2/M_{12}}$. This type of transformation is chosen
such that the coordinates are ``rationalized''\cite{mcguire}.
In terms of the new variables the Hamiltonian is written as:
\begin{align*}
   H_0 &= \frac{1}{2} \left(\mathbf{r}^2 -
     {\mbox{\boldmath $\nabla$}_\mathbf{r}}^2\right) \\
   V &= \,
   \Bigg[\frac{g_{12}}{\sigma}  \frac{\mu_{12}}{\sqrt{\mu_{123}}} \delta \! \left( x \right) +
   \frac{g_{23}}{\sigma}\frac{\sqrt{\mu_{123}}}{\mu_{12}} \delta \! \left(\frac{\mu_{123}}{m_1}x + y\right)\\
    &\hspace{4mm}+ \frac{g_{31}}{\sigma}\frac{\sqrt{\mu_{123}}}{\mu_{12}} \delta \! \left(-\frac{\mu_{123}}{m_2}x + y \right) \Bigg] \; .
\end{align*}
We once more notice that the full Hamiltonian is separable in terms of
center-of-mass motion and the relative motion just as in the case of
equal masses. For the relative motion, which happens in the $xy$-plane
we again use the (Jacobi) hyperspherical coordinates given by $\rho =
\sqrt{x^2+y^2}$, $\rho\in[0,\infty[$, and $\tan\phi=y/x$,
$\phi\in[-\pi,\pi[$.

\begin{figure}
\centering
\includegraphics[width=\columnwidth]{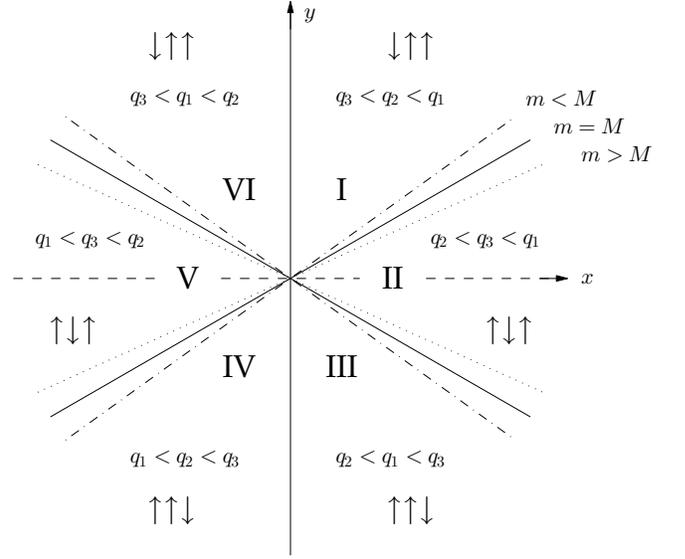}
\caption{Same as figure \ref{fig:hypersfaerisk} for the
  mass-imbalanced case. Note that the solid lines where $m=M$ form a
  $30^\circ$ angle with the $x$-axis. If $m>M$ the angle decreases --
  shown by the dotted line and the angle goes to zero as $m \gg M$. On
  the other hand if $m<M$ the angle increases -- shown by the dashed
  line and the angle goes to $45^\circ$ as $M\gg m$. This changes the
  shape and size of the domains in which the function $f(\mu,\phi)$
  lives.}
\label{particlepos}
\end{figure}

Our case of interest is a system of two identical fermions and a third
particle described with the following set of parameters: $g_{12}=0$, $g_{23}=g_{31}=g$, $m_1=m_2=m$ and $m_3=M$. For this case
we write the Hamiltonian in terms of the hyperspherical coordinates:
\begin{align*}
H &= \frac{1}{2} \left(z^2 - \frac{\partial^2}{\partial z^2}
  + \rho^2 -\frac{1}{\rho}\frac{\partial}{\partial \rho}
  - \frac{\partial^2}{\partial \rho^2}
  - \frac{1}{\rho^2} \frac{\partial^2}{\partial \phi^2} \right) \\[0.5em]
& + \frac{g}{\sigma\rho}\sqrt{\frac{2\gamma}{\gamma^2+1}} \sum_\pm \Big( \delta\big(\phi\pm\theta_0\big)+\delta\big(\phi\pm\theta_0-\pi\big)\Big) \; ,
\end{align*}
where $\gamma \equiv \frac{\mu_{123}}{m} = \sqrt{\frac{1}{1+2m/M}}$
and $\theta_0=\arctan \gamma$ is the angle between the $x$-axis and
the $q_2=q_3$ line. Notice that $\theta_0$ only depends on the mass
ratio, thus the mass ratio determines the size of each domain in
configuration space, seen on figure~\ref{particlepos}. The wave
functions at the $g=0$ and $1/g=0$ limits will again be products of
\eqref{eq:cm_wf} and \eqref{eq:relative_wf}, but the conditions for
finding the right angular functions $f(\mu,\phi)$ changes as
$\theta_0$ is not necessarily $\tfrac{\pi}{6}$. Let us follow the same
scheme as in Section \ref{sec:limits} for finding the angular
functions in the weakly and strongly interacting limits.  First we
note that the wave function for the relative motion satisfies
\begin{enumerate}
\item  $ \psi(-x,y)=-\psi(x,y)$ (Pauli principle),
\item  $\psi(x,y) \mapsto \psi(-x,-y)=\pm\psi(x,y)$ (parity).
\end{enumerate}
Next we can derive $\delta$-boundary condition similar to
Eq. \eqref{eq:condition-G} with parameter $\beta$ instead of $G$,
$\beta \equiv \frac{2\rho g}{\sigma}
\sqrt{\frac{2\gamma}{\gamma^2+1}}$. Notice that $\beta$ is
proportional to $G$ when the masses are fixed.

As before we start our analysis by treating $\beta$ as a constant for
all hyperradii ($\rho$) which is an effective 'toy model' of the
system. This schematic toy model employed in previous sections was
introduced for equal masses in Ref.~\cite{lindgren2014} and here we
generalize this model to mass imbalanced 2+1 systems. As we saw in the
previous section for the case of equal masses, the toy model
accurately reproduces the shape of the energy spectrum of the initial
Hamiltonian.  After applying the conditions and solving it, one can
show that for the \emph{odd parity} solutions we have
\begin{align}
  \label{eq:massimb_odd}
\mu \cos\left(\mu\tfrac{\pi}{2}\right) + \beta \sin\left(\mu (\tfrac{\pi}{2}-\theta_0)\right) \cos(\mu \theta_0) = 0 \; .
\end{align}
 In the same way for \emph{even
  parity} solutions parameter $\mu$ satisfies the following equation
\begin{align}
  \label{eq:massimb_even}
\mu \sin\left(\mu \tfrac{\pi}{2}\right) + \beta  \sin\left(\mu (\tfrac{\pi}{2}-\theta_0)\right) \sin\left(\mu \theta_0\right) = 0 \; .
\end{align}
Notice that when $m=M$ then $\gamma = \tfrac{1}{\sqrt{3}}$ yielding
$\theta_0=\tfrac{\pi}{6}$, and hence we have the same result as
calculated before.

\begin{figure*}[tbp]
\centering
\includegraphics[width=2\columnwidth]{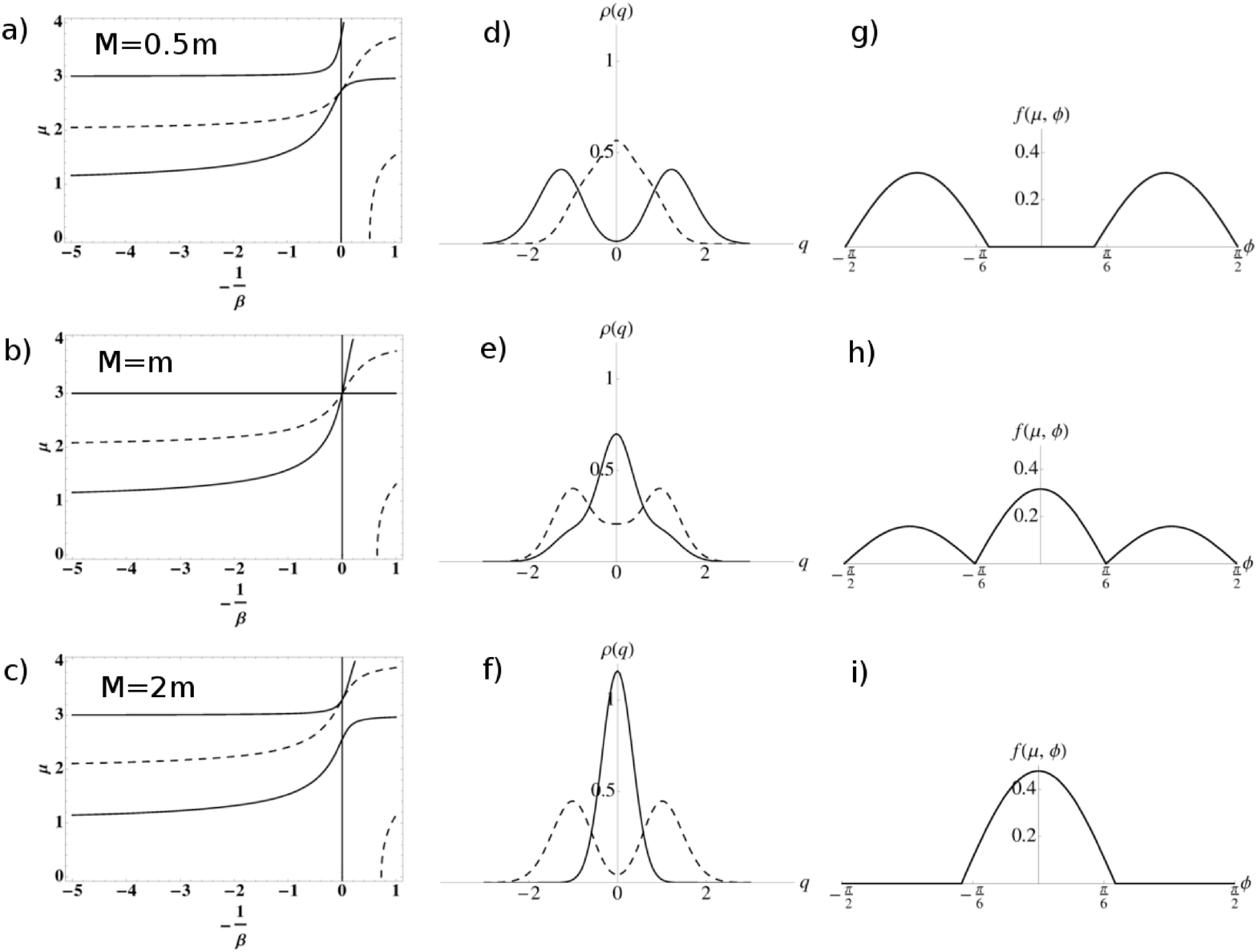}
\caption{The toy model energy spectra (solutions for $\mu$, neglecting
  the constant off-set energy) are shown in (a), (b) and (c) for
  different mass ratios. Solid lines denote odd parity solutions, and
  dashed lines denote even parity solutions. Notice that for $\beta>0$
  (thus $g > 0$) the ground state wave function is always odd in
  parity (solid). The ground state particle densities in the strongly
  interacting limit for the same mass ratios are shown in (d), (e) and
  (f). The solid line is for the single particle subsystem and the dashed
  line is for the 2-particle subsystem. Both are normalized to
  one. The angular part of the corresponding wave functions are shown
  for $\phi\in [-\tfrac{\pi}{2}, \tfrac{\pi}{2}]$ in (g), (h) and
  (i). Notice how the size of the domains I, II and III changes.}
\label{fig2}
\end{figure*}

Fig.~\ref{fig2}a, Fig.~\ref{fig2}b and Fig.~\ref{fig2}c show the
energy spectrum (solutions for $\mu$ to Eqs. \eqref{eq:massimb_odd}
and \eqref{eq:massimb_even}) with different mass ratios. Notice how
the horizontal-line solutions for odd parity for $m=M$ vanishes
instantaneously when the mass difference is different than $1$. Also
the ground state for $g>0$ remains odd in parity for any mass ratio,
but the degeneracy $1/g=0$ changes.

The toy model gives us knowledge of $\mu$ at $1/g=0$ so now we can
find the ground state wave functions for strongly interacting
systems. It turns out that only when $M=m$, we can construct a ground
state\footnote{It is worth to note that for the excited states there
  might be mass ratios that allow one to construct a wave function
  that is non-zero in all domains.} whose angular part can exist in
all domains (I, II, III, IV, V and VI), whenever $M<m$ the wave
function vanishes on II and V domains and whenever $M>m$ the wave
function must vanish on I, III, IV and VI domains. This happens
because for $m\neq M$ the spatial areas of I and II domains are
different and the ground state should live on the domain with the
largest area. This means that the number of allowed domains for
angular part is reduced instantaneously for any small mass
imbalance. Notice that for $M<m$ the wave function is double
degenerate since there are 4 allowed domains and for $M>m$ we have a
single degeneracy. We illustrate this discussion with examples for
$M=m$, $M<m$ and $M>m$. We can find the exact wave functions at
$1/g=0$ in the same way as we did in section \ref{sec:impenetrable}.
The angular part of the wave function for the ground state with $M=m$
is found to be:
\begin{equation*}
f_{1}^{\infty}(3,\phi)
= \begin{cases}
-\cos(3\phi) & \text{in I}\\
2 \cos(3\phi) & \text{in II}\\
-\cos(3\phi) & \text{in III}
\end{cases}
\end{equation*}
where domains I, II and III are separated by the solid lines in
Fig.~\ref{particlepos}. When $M<m$, $\mu$ is no longer an integer. For
instance, when $M=\tfrac{1}{2}m$, $\theta_0=0.421$ (or $24.1^\circ$)
and $\mu=2\pi/[\pi-\theta_0(\text{rad})] \approx 2.731$ hence the
wave function is
\begin{equation*}
f_{1}^{\infty}(\mu,\phi)
= \begin{cases}
\sin(\mu(\phi-\theta_0)) & \text{in I}\\
0 & \text{in II}\\
-\sin(\mu(\phi+\theta_0)) & \text{in III}
\end{cases}
\end{equation*}
Notice that the boundaries of domains change also: for this case the
domains are separated by the dotted lines in Fig.~\ref{particlepos}.
As an example of $M>m$, we take $M=2m$. In this case $\theta_0=0.615$
(or $35.2^\circ$) and $\mu=\pi/[2\theta_0(\text{rad})] \approx 2.552$
and the wave function can then be constructed as
\begin{equation*}
f_{1}^{\infty}(\mu,\phi)
= \begin{cases}
0 & \text{in I}\\
\cos(\mu\phi) & \text{in II}\\
0 & \text{in III}
\end{cases}
\end{equation*}
The wave functions along with the corresponding densities calculated
just like in section \ref{sec:density} are illustrated in
Fig.~\ref{fig2}. One might think that there would be some continuous
crossover from $M<m$ to $M=m$ and then to $M>m$ but this is not the
case. As shown in Fig.~\ref{fig2}(a) the case with $M=m$ generates
almost a singularity point where three states with two states of the
same parity cross one another at $1/g=0$. However, when $M$ is
slightly bigger or smaller than $m$, this threefold degeneracy is
lifted instantly and the ground state wave function is non-zero only
in certain domains.

The results presented in this section allows us to elucidate the exact
behavior of the system in the strongly interaction limit. When $M>m$
we have $30^\circ<\theta_0<45^\circ$ and domain II is favored for the
ground state ($\uparrow\downarrow\uparrow$ if we again think of
spin-$\tfrac{1}{2}$ particles), while for $M<m$ we have
$0<\theta_0<30^\circ$ where domain I and III are then favored
($\uparrow\uparrow\downarrow$ and $\downarrow\uparrow\uparrow$).  The
special case with $M=m$ has $\theta_0=30^\circ$ and the wavefunction
for the ground state is spread over all regions.  It is important to
notice that there is no continuous crossover going from $M<m$ to $M=m$
and then to $M>m$ for strongly interacting systems, i.e.  any small
infinitesimal mass imbalanced requires the wave function to vanish at
a certain region and the 'accidental' degeneracy of the spectrum at
$M=m$ is immediately broken.

\section{Conclusion}
\label{sec:conclusions}
We studied a quantum mechanical system of three particles confined to
a one-dimensional harmonic trap potential consisting of two
indistinguishable fermions interacting with a third particle via a
zero-range contact interaction of strength $g$. We showed how the
Schr\"odinger equation is solved in the limits of no interaction,
$g=0$, and infinitely strong repulsion, $1/g=0$. Then we assumed a
factorized form of the wave function for intermediate values of $0 < g
< \infty$ and provided a class of approximated wave functions that
reduce to the analytic solutions in the limits $g\rightarrow 0$ and $g
\rightarrow \infty$. We produced a basis of variational wave functions
for every value of $g$, and we found that the resulting energy
spectrum is very close to the numerically calculated energies. A
diagonalization of the Hamiltonian in a basis of 54 approximated wave
functions yielded even better results, as expected.

Furthermore, we calculated the probability densities of the
approximated wave functions for the ground and first excited state,
and discussed ferro- and antiferromagnetic behavior. In order to
discuss this in the language of spin algebra, we took the three
particles to be spin-$\tfrac{1}{2}$ particles with the two
indistinguishable particles being the spin-up state, while the third
particle was in the spin-down state.

Finally, we studied the case where the mass of the indistinguishable
particles was allowed to differ from the mass of the third
particle. This was done in a schematic 'toy model' where the coupling
is re-scaled with the hyperradius (effectively factorizing the
problem).  This model becomes exact in the non-interacting and
strongly interacting limits. Here we find a most interesting behavior
of the degeneracy of the ground state at infinite coupling
strength. When the impurity is heavier than the two identical
particles, we obtain a non-degenerate ground state, while in the
opposite case we find a doubly degenerate ground states. The doubly
degenerate ground state is also seen in the two-component bosonic
system with equal masses discussed in Ref.~\cite{dehk2014}.  In this
sense, the equal mass case is very special with its triply degenerate
ground state for strong interaction.  But given that Nature provides
us with numerous two-component systems of equal mass this is of course
an extremely important special case at that.

This research was supported by the Danish Council for Independent 
Research DFF Natural Sciences and the DFF Sapere Aude program.


\onecolumngrid 
\appendix


\section{Equations for $\mu$}
\label{sec:solvedmu}

From \eqref{eq:condition-pauli}
\begin{equation*}
  f\left(\mu,\tfrac{\pi}{2}\right) =
  A \, \cos\left(\mu\tfrac{\pi}{2}\right) +
  B \, \sin\left(\mu\tfrac{\pi}{2}\right) = 0 \; .
\end{equation*}
Considering only the states with even parity we get from
\eqref{eq:condition-parity} that
\begin{equation*}
  C \, \cos(-\mu\phi) + D \, \sin(-\mu\phi)
  = -C \, \cos(\mu\phi) - D \, \sin(\mu\phi)
 \Rightarrow C = 0 \; .
\end{equation*}
By the continuity at $\phi = \tfrac{\pi}{6}$
\begin{equation*}
  A \, \cos\left(\mu\tfrac{\pi}{6}\right) +
  B \, \sin\left(\mu\tfrac{\pi}{6}\right)
  = D \, \sin\left(\mu\tfrac{\pi}{6}\right) \; ,
\end{equation*}
and finally from \eqref{eq:condition-G}
\begin{equation*}
  -\mu A \, \sin\left(\mu\tfrac{\pi}{6}\right) +
  \mu B \, \cos\left(\mu\tfrac{\pi}{6}\right)
  -\mu D \, \cos\left(\mu\tfrac{\pi}{6}\right)
  = G D \, \sin\left(\mu\tfrac{\pi}{6}\right) \; .
\end{equation*}
Collecting these equations in a matrix equation yields
\begin{equation*}
  \begin{bmatrix} 0 &
    \cos\left(\mu\tfrac{\pi}{2}\right) &
    \sin\left(\mu\tfrac{\pi}{2}\right) \\[0.5em]
    -\sin\left(\mu\tfrac{\pi}{6}\right) &
    \cos\left(\mu\tfrac{\pi}{6}\right)&
    \sin\left(\mu\tfrac{\pi}{6}\right) \\[0.5em]
    -\cos\left(\mu\tfrac{\pi}{6}\right) -
    \tfrac{G}{\mu} \, \sin\left(\mu\tfrac{\pi}{6}\right) &
    -\sin\left(\mu\tfrac{\pi}{6}\right) &
    \cos\left(\mu\tfrac{\pi}{6}\right)
  \end{bmatrix}
    \begin{bmatrix} D \\[0.5em] A \\[0.5em] B \end{bmatrix}
    = \begin{bmatrix} 0 \\[0.5em] 0 \\[0.5em] 0 \end{bmatrix} \; ,
\end{equation*}
which have a solution only if the determinant of the matrix is
zero. Doing the the same calculation for odd parity, and setting the
determinants of the two matrices to zero, we arrive at the equations
\eqref{eq:determinant-even} and \eqref{eq:determinant-odd}.


\section{Approximated angular functions}
\label{sec:angular}

We would like to find an expression for the angular part of the
approximated wave function for the interacting states that have the
right parity and reduces to the known solutions in the limits $g
\rightarrow 0$ and $g \rightarrow \infty$. First, we consider a state
with odd parity ($\mu_0$ is odd), and $f_{\mu_0}^g(\mu,\phi)$ must
therefore be symmetric around $\phi = 0$. Also it must be zero when
$\phi = \pm \tfrac{\pi}{2}$, and so we make the ansatz
\begin{equation*}
  f_{\mu_0}^g (\mu,\phi) =
   \begin{cases}
     \sin\left(\mu\left(\tfrac{\pi}{2}-\phi\right)\right)
     \quad &\text{in I}\\[0.5em]
     A + B\,\cos(\mu\phi)
     \quad &\text{in II}\\[0.5em]
     \sin\left(\mu\left(\tfrac{\pi}{2}+\phi\right)\right)
     \quad &\text{in III}
   \end{cases}
\end{equation*}
Since we assume the factorized form \eqref{eq:assume_wf} for
the wave function for the relative motion, we indirectly assume that
$G = \sqrt{2} \, g \,\rho$ is independent of $\rho$, and we apply
\eqref{eq:condition-G} at $\phi_0 = \tfrac{\pi}{6}$:
\begin{align*}
  \Delta\left(\left.\frac{\partial f_{\mu_0}^g(\mu,\phi)}{\partial\phi}\right|_{\tfrac{\pi}{6}}\right)
  &= -\mu\,\cos\left(\mu\tfrac{\pi}{3}\right)
  + \mu B \,\sin\left(\mu\tfrac{\pi}{6}\right) \\
  &=G\,\sin\left(\mu\tfrac{\pi}{3}\right)
\end{align*}
Isolating $G$ from \eqref{eq:determinant-odd} and inserting in the
above yields
\begin{align*}
  -\mu\,\cos\left(\mu\tfrac{\pi}{3}\right) + \mu B \,\sin\left(\mu\tfrac{\pi}{6}\right)
  = -\mu \, \frac{\cos\left(\mu\tfrac{\pi}{2}\right)}
  {\cos\left(\mu\tfrac{\pi}{6}\right)} \; ,
\end{align*}
and then we can find the constant $B$ as
\begin{equation*}
  B = \frac{1}{\sin\left(\mu\tfrac{\pi}{6}\right)} \left[
      \cos\left(\mu\tfrac{\pi}{3}\right)
      - \frac{\cos\left(\mu\tfrac{\pi}{2}\right)}{\cos\left(\mu\tfrac{\pi}{6}\right)}
      \right] = 2 \, \sin\left(\mu\tfrac{\pi}{6}\right) \; .
\end{equation*}
The function must be continuous at $\phi = \tfrac{\pi}{6}$:
\begin{equation*}
  A + B \, \cos\left(\mu\tfrac{\pi}{6}\right)
  = \sin\left(\mu\tfrac{\pi}{3}\right) \; ,
\end{equation*}
and then $A$ is found as
\begin{equation*}
  A = \sin\left(\mu\tfrac{\pi}{3}\right) - B \, \cos\left(\mu\tfrac{\pi}{6}\right)
  = 0 \; .
\end{equation*}
The angular function in domain II can be written as
\begin{equation*}
   A + B\,\cos(\mu\phi)
   = \sin\left(\mu\left(\tfrac{\pi}{6}-\phi\right)\right)
   + \sin\left(\mu\left(\tfrac{\pi}{6}+\phi\right)\right) \; ,
\end{equation*}
and thus in total we get
\begin{align*}
  f_{\mu_0}^g (\mu,\phi) =
   \begin{cases}
     \sin\left(\mu\left(\tfrac{\pi}{2}-\phi\right)\right)  &\text{in I}\\[0.5em]
     \sin\left(\mu\left(\tfrac{\pi}{6}-\phi\right)\right)
   + \sin\left(\mu\left(\tfrac{\pi}{6}+\phi\right)\right)  &\text{in II}\\[0.5em]
     \sin\left(\mu\left(\tfrac{\pi}{2}+\phi\right)\right)  &\text{in III}
   \end{cases}
\end{align*}
One may check that it reduces to the known solutions in the
interaction limits, for instance we recover the ground state in the
limits by putting $\mu = 1$ or $\mu = 3$.

We can do the same analysis for a state with even parity ($\mu_0$ is
even), and we arrive at the following angular function:
\begin{align*}
  f_{\mu_0}^g (\mu,\phi) =
   \begin{cases}
     \sin\left(\mu\left(\tfrac{\pi}{2}-\phi\right)\right)   &\text{in I}\\[0.5em]
     \sin\left(\mu\left(-\tfrac{\pi}{6}+\phi\right)\right)
   + \sin\left(\mu\left(\tfrac{\pi}{6}+\phi\right)\right)   &\text{in II}\\[0.5em]
     \sin\left(\mu\left(-\tfrac{\pi}{2}-\phi\right)\right)  &\text{in III}
   \end{cases}
\end{align*}
Notice that angular function for even parity is just the same as for
odd parity with some signs reversed. Thus the general angular function
for the interacting states can be expressed as \eqref{eq:angular} in
the main text.


\section{The Hamiltonian matrix elements}
\label{sec:energy}

We want to calculate the matrix elements for the Hamiltonian in the
basis of approximated wave functions. As noted in the main text, we
can consider three cases of a matrix element between interacting and
non-interacting states, one of which (two non-interacting states) is
simply \eqref{eq:matrixel-nonnon}. We now consider the case of a
matrix element between an interacting and a non-interacting
state. Since the Hamiltonian matrix is symmetric, we can take $H$ to
act on the non-interacting state in which case we get
\begin{align*}
  &\tensor*[^{\text{ap}}_g]
  {\left<\nu',\mu_0',\eta'\vphantom{H} \right| H
    \left| \vphantom{H \mu_0'} \nu,\mu_0,\eta \right>}{} \\
  &=
  \tensor*[^{\text{ap}}_g]
  {\left<\nu',\mu_0',\eta'\vphantom{H_0} \right| H_0
    \left| \vphantom{H_0 \mu_0'} \nu,\mu_0,\eta \right>}{} \\
  &=
  \left(\tfrac{3}{2} + 2\nu + \mu_0 + \eta\right) \,
  \tensor*[^{\text{ap}}_g]
  {\left<\nu',\mu_0',\eta'\, \right| \left. \! \vphantom{\mu_0'}
      \nu,\mu_0,\eta \right>}{} \\
  &=
  \left(\tfrac{3}{2} + 2\nu + \mu_0 + \eta\right) \, \delta_{\eta'\eta} \,
  \int\limits_0^\infty \mathrm{d} \rho \, \rho \,
  R_{\nu'} (\mu',\rho) R_\nu (\mu,\rho)
  \int\limits_{-\pi}^\pi \mathrm{d} \phi \,
  f_{\mu_0'}^g (\mu',\phi) f_{\mu_0}^g (\mu,\phi) \; ,
\end{align*}
With the non-interacting angular function $f_{\mu_0}^g(\mu,\phi) =
\cos(\mu\phi)$ or $f_{\mu_0}^g(\mu,\phi) = \sin(\mu\phi)$ with $\mu
\equiv_3 0$ and $f_{\mu_0'}^g(\mu',\phi)$ given by \eqref{eq:angular},
one may verify that
\begin{align*}
  \int\limits_{-\pi}^\pi \mathrm{d} \phi \,
  f_{\mu_0'}^g (\mu',\phi) f_{\mu_0}^g (\mu,\phi) = 0 \; .
\end{align*}
We now turn to the matrix element between two interacting states. We
calculate the $H_0$ term and $V$ term separately, starting with the
former using \eqref{eq:H0_cyl}. If the double derivative of
$f_{\mu_0}^g (\mu,\phi)$ with respect to $\phi$ was defined for every
$\phi \in [-\pi,\pi[$, this would be straightforward. However, this is
not the case since it is not continuously differentiable in $\phi =
\pm \tfrac{\pi}{6}$ and $\phi = \pm \tfrac{5\pi}{6}$ when $g \neq 0$,
so we isolate this part of the matrix element and treat it carefully.
\begin{align*}
  &\tensor*[^{\text{ap}}_g]
  {\left<\nu',\mu_0',\eta'\vphantom{H_0} \right| H_0
    \left| \vphantom{H_0 \mu_0'} \nu,\mu_0,\eta \right>}
  {^{\text{ap}}_g} \\
  &=
  \int\limits_{-\infty}^\infty \mathrm{d} z
  \int\limits_0^\infty \mathrm{d} \rho \, \rho
  \int\limits_{-\pi}^\pi \mathrm{d} \phi \,
  \psi_{\eta'}(z) R_{\nu'}(\mu',\rho) f_{\mu_0'}^g(\mu',\phi)
  \frac{1}{2} \left(z^2 - \frac{\partial^2}{\partial z^2}
  + \rho^2 -\frac{1}{\rho}\frac{\partial}{\partial\rho}
  - \frac{\partial^2}{\partial \rho^2}
      - \frac{1}{\rho^2} \frac{\partial^2}{\partial \phi^2} \right)
  \psi_{\eta}(z) R_{\nu}(\mu,\rho) f_{\mu_0}^g(\mu,\phi) \\
  &=
  \int\limits_{-\infty}^\infty \mathrm{d} z
  \int\limits_0^\infty \mathrm{d} \rho \, \rho
  \int\limits_{-\pi}^\pi \mathrm{d} \phi \,
  \psi_{\eta'}(z) R_{\nu'}(\mu',\rho) f_{\mu_0'}^g(\mu',\phi)
  \frac{1}{2} \left(z^2 - \frac{\partial^2}{\partial z^2}
  + \rho^2 -\frac{1}{\rho}\frac{\partial}{\partial\rho}
  - \frac{\partial^2}{\partial \rho^2}
      + \frac{\mu^2}{\rho^2} \right)
  \psi_{\eta}(z) R_{\nu}(\mu,\rho) f_{\mu_0}^g(\mu,\phi) \\
  &\hphantom{=} - \frac{\mu^2}{2} \delta_{\eta',\eta}
  \int\limits_0^\infty \mathrm{d} \rho \, \rho^{-1}
  R_{\nu'}(\mu',\rho) R_{\nu}(\mu,\rho)
  \int\limits_{-\pi}^\pi \mathrm{d} \phi \,
  f_{\mu_0'}^g(\mu',\phi) f_{\mu_0}^g(\mu,\phi) \\
  &\hphantom{=} - \frac{1}{2} \delta_{\eta',\eta}
  \int\limits_0^\infty \mathrm{d} \rho \, \rho^{-1}
  R_{\nu'}(\mu',\rho) R_{\nu}(\mu,\rho)
  \int\limits_{-\pi}^\pi \mathrm{d} \phi \,
  f_{\mu_0'}^g(\mu',\phi) \frac{\partial^2}{\partial \phi^2} f_{\mu_0}^g(\mu,\phi)
\end{align*}
We now treat this last integral over $\phi$. Since the integrand is
not always defined, we must divide the integral into pieces and take
the limit as we integrate over the problematic points and over the
intervals between them.
\begin{align*}
  &\int\limits_{-\pi}^\pi \mathrm{d} \phi \, f_{\mu_0'}^g (\mu',\phi)
     \frac{\partial^2}{\partial \phi^2} \, f_{\mu_0}^g(\mu,\phi)
     = \lim\limits_{\varepsilon \rightarrow 0} \left(
       \int\limits_{-\frac{\pi}{6}-\varepsilon}^{-\frac{\pi}{6}+\varepsilon}
       \mathrm{d} \phi \, f_{\mu_0'}^g (\mu',\phi)
     \frac{\partial^2}{\partial \phi^2} \, f_{\mu_0}^g(\mu,\phi) \right. + \\[0.3em]
     &\left.
     \int\limits_{-\frac{\pi}{6}+\varepsilon}^{\frac{\pi}{6}-\varepsilon}
     \mathrm{d} \phi \, f_{\mu_0'}^g (\mu',\phi)
     \frac{\partial^2}{\partial \phi^2} \, f_{\mu_0}^g(\mu,\phi)
     + \ldots +
     \int\limits_{-\frac{5\pi}{6}+\varepsilon}^{-\frac{\pi}{6}-\varepsilon}
     \mathrm{d} \phi \, f_{\mu_0'}^g (\mu',\phi)
     \frac{\partial^2}{\partial \phi^2} \, f_{\mu_0}^g(\mu,\phi)
     \right) \; .
\end{align*}
Consider a $\phi_0 \in \big\{ \pm \tfrac{\pi}{6}, \pm \tfrac{5\pi}{6}
\big\}$ and the integral over this point. Integrals of that type will
in the limit $\varepsilon \rightarrow 0$ be evaluated as
\begin{align*}
  \lim\limits_{\varepsilon\rightarrow 0}
  \int\limits_{\phi_0-\varepsilon}^{\phi_0+\varepsilon} \mathrm{d} \phi \,
  f_{\mu_0'}^g (\mu',\phi) \frac{\partial^2}{\partial \phi^2} \, f_{\mu_0}^g(\mu,\phi)
  &=
  f_{\mu_0'}^g (\mu',\phi_0) \,
  \lim\limits_{\varepsilon\rightarrow 0}
  \left(\left. \frac{\partial f_{\mu_0}^g}{\partial\phi} \right|_{\phi_0+\varepsilon}
  - \left. \frac{\partial f_{\mu_0}^g}{\partial \phi} \right|_{\phi_0-\varepsilon} \right) \\
  &=
  f_{\mu_0'}^g (\mu',\phi_0) \,
  \Delta\left(\left. \frac{\partial f_{\mu_0}^g}{\partial\phi} \right|_{\phi_0}\right) \; .
\end{align*}
It follows from the antisymmetry $f_{\mu_0}^g (\mu,-x) = - f_{\mu_0}^g
(\mu,x)$ that we get the same contribution from the $x>0$ and $x<0$
half planes, and so we only need to compute it for $\phi_0 = \pm
\tfrac{\pi}{6}$. If we plug in the angular function
\eqref{eq:angular}, it is easy to show that
\begin{align*}
  f_{\mu_0'}^g (\mu',\tfrac{\pi}{6}) \,
  \Delta\left(\left. \frac{\partial f_{\mu_0}^g}{\partial \phi} \right|_{\tfrac{\pi}{6}}\right) +
  f_{\mu_0'}^g (\mu',-\tfrac{\pi}{6}) \,
  \Delta\left(\left. \frac{\partial f_{\mu_0}^g}{\partial \phi} \right|_{-\tfrac{\pi}{6}}\right)
  =
  - \left(1+(-1)^{\mu_0'+\mu_0}\right) \mu \sin\left(\mu'\tfrac{\pi}{3}\right)
  \left(2\cos\left(\mu\tfrac{\pi}{3}\right) + (-1)^{\mu_0}\right)
\end{align*}
On the intervals between the points where the double derivative is
undefined, we have a well-defined second derivative
$\frac{\partial^2}{\partial \phi^2}f_{\mu_0}^g(\mu,\phi) = -\mu^2
f_{\mu_0}^g(\mu,\phi)$. The sum of integrals over these intervals have
the limit
\begin{align*}
  \lim\limits_{\varepsilon\rightarrow 0} \left(
  \int\limits_{-\tfrac{\pi}{6}+\varepsilon}^{\tfrac{\pi}{6}-\varepsilon}
  \mathrm{d} \phi \,
  f_{\mu_0'}^g (\mu',\phi) \frac{\partial^2}{\partial \phi^2} \, f_{\mu_0}^g(\mu,\phi) +
  \ldots +
  \int\limits_{\tfrac{5\pi}{6}+\varepsilon}^{-\tfrac{\pi}{6}-\varepsilon}
  \mathrm{d} \phi \,
  f_{\mu_0'}^g (\mu',\phi) \frac{\partial^2}{\partial \phi^2} \, f_{\mu_0}^g(\mu,\phi)
  \right)
  = -\mu^2
  \int\limits_{-\pi}^\pi \mathrm{d} \phi \,
  f_{\mu_0'}^g(\mu',\phi) f_{\mu_0}^g(\mu,\phi) \; .
\end{align*}
Thus
\begin{align*}
  &\int\limits_{-\pi}^\pi \mathrm{d} \phi \,
  f_{\mu_0'}^g(\mu',\phi) \frac{\partial^2}{\partial \phi^2} f_{\mu_0}^g(\mu,\phi) \\
  &= - 2\left(1+(-1)^{\mu_0'+\mu_0}\right) \mu \sin\left(\mu'\tfrac{\pi}{3}\right)
  \left(2\cos\left(\mu\tfrac{\pi}{3}\right) + (-1)^{\mu_0}\right)
  -\mu^2
  \int\limits_{-\pi}^\pi \mathrm{d} \phi \,
  f_{\mu_0'}^g(\mu',\phi) f_{\mu_0}^g(\mu,\phi) \; .  
\end{align*}
Using that
\begin{align*}
  \frac{1}{2} \left(z^2 - \frac{\partial^2}{\partial z^2}
  + \rho^2 -\frac{1}{\rho}\frac{\partial}{\partial \rho} -
  \frac{\partial^2}{\partial \rho^2}
      + \frac{\mu^2}{\rho^2} \right)
  \psi_{\eta}(z) R_{\nu}(\mu,\rho) f_{\mu_0}^g(\mu,\phi)
  =
  \left(\tfrac{3}{2} + 2\nu + \mu + \eta\right)
  \psi_{\eta}(z) R_{\nu}(\mu,\rho) f_{\mu_0}^g(\mu,\phi)
\end{align*}
and the orthogonality of $\psi_{\eta}(z)$'s, it is now straightforward
to rewrite the matrix element on the form \eqref{eq:H0-intint}.

For the interaction term, we use \eqref{eq:V_cyl} and evaluate
\eqref{eq:angular} at the specified points, yielding
\begin{align*}
  &\tensor*[^{\text{ap}}_g]
    {\left<\nu',\mu_0',\eta' \vphantom{V\mu_0'} \right| V \left|
        \vphantom{V\mu_0'} \nu,\mu_0,\eta \right>}
  {^{\text{ap}}_g} \\
  &=
  \int\limits_{-\infty}^\infty \mathrm{d} z
  \int\limits_0^\infty \mathrm{d} \rho \, \rho
  \int\limits_{-\pi}^\pi \mathrm{d} \phi \,
  \psi_{\eta'}(z) R_{\nu'}(\mu',\rho) f_{\mu_0'}^g(\mu',\phi) \\
  &\hphantom{=} \times
  \frac{g}{\sqrt{2}\,\rho} \Big[
    \delta\! \left(\phi - \tfrac{\pi}{6}\right)
  + \delta\! \left(\phi + \tfrac{5\pi}{6}\right)
  + \delta\! \left(\phi + \tfrac{\pi}{6}\right)
  + \delta\! \left(\phi - \tfrac{5\pi}{6}\right)
    \Big]
  \psi_{\eta}(z) R_{\nu}(\mu,\rho) f_{\mu_0}^g(\mu,\phi) \\
  &=
  \frac{g}{\sqrt{2}} \delta_{\eta',\eta}
  \int\limits_0^\infty \mathrm{d} \rho \,
  R_{\nu'}(\mu',\rho) R_{\nu}(\mu,\rho)
  \left( 2 \sin(\mu'\tfrac{\pi}{3}) \sin(\mu\tfrac{\pi}{3})
  + 2 (-1)^{\mu_0' + \mu_0}
  \sin(\mu'\tfrac{\pi}{3}) \sin(\mu\tfrac{\pi}{3})\right)
\end{align*}
from which \eqref{eq:V-intint} follows immediately.

The matrix elements for $ H_0$ and $ V$ for the ground state
$\left|0,1,0\right>_g^\text{ap}$ and first excited state
$\left|0,2,0\right>_g^\text{ap}$ are found to be
\begin{align*}
  &\tensor*[^{\text{ap}}_g]
  {\left<0,1,0\vphantom{H_0} \left| H_0 \right|
      \vphantom{H_0}0,1,0\right>}
  {^{\text{ap}}_g}
  =
  \frac{3}{2} + \mu +
  \frac{\Gamma(\mu)}{\Gamma(\mu+1)}
  \frac{\mu^2 \sin\left(\frac{\pi}{3}\mu\right)
  \left(2 \cos\left(\frac{\pi}{3}\mu\right) - 1\right)}
  {\tfrac{2\pi}{3}\mu - \tfrac{\pi}{3}\mu\cos\left(\frac{\pi}{3}\mu\right)
  +\sin\left(\frac{\pi}{3}\mu\right)
  -\sin\left(\frac{2\pi}{3}\mu\right)} \\[1em]
  &\tensor*[^{\text{ap}}_g]
  {\left<0,1,0\vphantom{V} \left| V \right|
      \vphantom{V}0,1,0\right>}
  {^{\text{ap}}_g}
  =
  g \,
  \frac{\Gamma\big(\mu+\tfrac{1}{2}\big)}{\Gamma(\mu+1)}
  \frac{\sqrt{2} \mu \sin^2\left(\frac{\pi}{3}\mu\right)}
  {\tfrac{2\pi}{3}\mu - \tfrac{\pi}{3}\mu\cos\left(\frac{\pi}{3}\mu\right)
  +\sin\left(\frac{\pi}{3}\mu\right)
  -\sin\left(\frac{2\pi}{3}\mu\right)} \\[1em]
  &\tensor*[^{\text{ap}}_g]
  {\left<0,2,0\vphantom{H_0} \left| H_0 \right|
      \vphantom{H_0}0,2,0\right>}
  {^{\text{ap}}_g}
  =
  \frac{3}{2} + \mu +
  \frac{\Gamma(\mu)}{\Gamma(\mu+1)}
  \frac{\mu^2 \sin\left(\frac{\pi}{3}\mu\right)
  \left(2 \cos\left(\frac{\pi}{3}\mu\right) + 1\right)}
  {\tfrac{2\pi}{3}\mu + \tfrac{\pi}{3}\mu\cos\left(\frac{\pi}{3}\mu\right)
  -\sin\left(\frac{\pi}{3}\mu\right)
  -\sin\left(\frac{2\pi}{3}\mu\right)} \\[1em]
  &\tensor*[^{\text{ap}}_g]
  {\left<0,2,0\vphantom{V} \left| V \right|
      \vphantom{V}0,2,0\right>}
  {^{\text{ap}}_g}
  =
  g \,
  \frac{\Gamma\big(\mu+\tfrac{1}{2}\big)}{\Gamma(\mu+1)}
  \frac{\sqrt{2} \mu \sin^2\left(\frac{\pi}{3}\mu\right)}
  {\tfrac{2\pi}{3}\mu + \tfrac{\pi}{3}\mu\cos\left(\frac{\pi}{3}\mu\right)
  -\sin\left(\frac{\pi}{3}\mu\right)
  -\sin\left(\frac{2\pi}{3}\mu\right)} \; .
\end{align*}


\twocolumngrid 

\end{document}